\documentclass[12pt,article]{amsart}
\usepackage[latin1]{inputenc}
\usepackage{tikz}
\usetikzlibrary{arrows.meta}
\usetikzlibrary{bending}
\usepackage{verbatim}
\usepackage{xcolor}
\usepackage[all,color]{xy}
\usepackage{dblfloatfix,afterpage,float,longtable,tensor,mathrsfs,import,stmaryrd,wasysym,amsmath,amssymb,amsthm,amscd,amsxtra,amsfonts,mathrsfs,graphicx,enumerate,bm,slashed,array,setspace,enumitem,minibox,booktabs,rotating,multirow,adjustbox,hhline,arydshln,placeins}
\usepackage{import,stmaryrd,wasysym,amsmath,amssymb,amsthm,amscd,amsxtra,amsfonts,mathrsfs,graphicx,enumerate,bm,slashed,array,setspace,enumitem,minibox,booktabs,rotating,multirow,adjustbox,hhline,arydshln,placeins}
\usepackage{tikz-feynman}
\input xy
\xyoption{all}
\usetikzlibrary{shapes,arrows}
\usetikzlibrary{positioning}
\usetikzlibrary{calc}
\input xy
\xyoption{all}
\newtheorem{Theorem}{Theorem}[section]

\newtheorem{Definition}[Theorem]{Definition}

\newtheorem*{Definition*}{Definition A}
\newtheorem*{Theorem*}{Theorem B}
\DeclareMathOperator{\im}{im}
\DeclareMathOperator{\vol}{vol}

\advance\evensidemargin-.5in
\advance\oddsidemargin-.5in
\advance\textwidth1in

\definecolor{darkgreen}{rgb}{0.0, 0.5, 0.0}
\definecolor{darkred}{rgb}{0.7, 0.11, 0.11}

\makeatletter
\DeclareRobustCommand{\cev}[1]{%
  {\mathpalette\do@cev{#1}}%
}
\newcommand{\do@cev}[2]{%
  \vbox{\offinterlineskip
    \sbox\z@{$\m@th#1 x$}%
    \ialign{##\cr
      \hidewidth\reflectbox{$\m@th#1\vec{}\mkern4mu$}\hidewidth\cr
      \noalign{\kern-\ht\z@}
      $\m@th#1#2$\cr
    }%
  }%
}
\makeatother

\begin{document}

\author{Charlie Beil}
\address{Institut f\"ur Mathematik und Wissenschaftliches Rechnen, Universit\"at Graz, Heinrichstrasse 36, 8010 Graz, Austria.}
 \email{charles.beil@uni-graz.at}
\title[A comb.\ derivation of the SM interactions from the Dirac Lagrangian]{A combinatorial derivation of the standard model interactions from the Dirac Lagrangian}

\begin{abstract}
A composite model of the standard model particles was recently derived using the Dirac Lagrangian on a spacetime where time does not advance along the worldlines of fundamental dust particles, called an `internal spacetime'.
The aim of internal spacetime geometry is to model certain quantum phenomena using (classical) \textit{degenerate} spacetime metrics.
For example, on an internal spacetime, tangent spaces have variable dimension, and spin wavefunction collapse is modeled by the projection from one tangent space to another.
In this article we show that the combinatorial structure of the internal Dirac Lagrangian yields precisely the standard model trivalent vertices, together with two additional new (longitudinal) $Z$ vertices that generate the four-valent boson vertices. 
In particular, we are able to derive electroweak parity violation for both leptons and quarks.
We also obtain new restrictions on the possible spin states that can occur in certain interactions.
Finally, we determine the trivalent vertices of the new massive spin-$2$ boson predicted by the model.
\end{abstract}

\maketitle

\tableofcontents

\section{Introduction}

A new framework of \textit{degenerate} spacetime metrics, based on a generalized notion of simultaneity, was recently introduced in \cite{B6} with the hope that it may resolve the quantum measurement problem and serve as an approximate geometric model of quantum phenomena. 
The basis of the framework is a simple modification to (classical) general relativity: time does not advance along the worldlines of fundamental dust particles. 
It was shown in \cite{B6} that such particles are charged and have spin $\tfrac 12$.
We call these fundamental particles \textit{pointons}, and the resulting geometry \textit{internal spacetime}, or simply \textit{spacetime}.

In \cite{B1} we investigated the Dirac Lagrangian $\mathcal{L}$ on an internal spacetime and found that the chirality of a pointon spinor (under $\gamma^5$) is electric charge. 
From this new identification, nearly the entire standard model, with three generations of quarks and leptons, the electroweak bosons, and the Higgs boson, all appear as combinatorial configurations of pointon spinors obtained from the mass term of $\mathcal{L}$; see Table \ref{particles}. 
These configurations, which we call \textit{geoms} for `geometric atoms', are bound states of at most four pointons that are constrained by the Dirac equation and the Pauli exclusion principle. 
The only remaining standard model particle is the gluon, which we model in \cite{B2} using a construction similar to flux tubes. 
Moreover, only one new particle arises that does not belong to the standard model, namely, a massive spin-$2$ boson.

\begin{figure}
\begin{center}
$\begin{array}{cc}
\xy  0;/r.18pc/:
(0,10)*{}="1"; (0,50)*{}="2";(0,20)*{}="3";
{\ar@{..>}|-{\text{\footnotesize{time}}}"3";"2"};
\endxy &
\includegraphics[width=4.3cm]{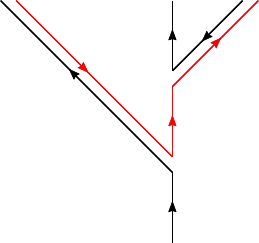}
\end{array}$
\end{center}
\caption{The single lines are electron worldlines, and the double lines (which are separated for clarity) are photon worldlines.}
\label{electron picture}
\end{figure}

In this article we consider interactions between geoms.
These interactions should correspond to the trivalent interaction vertices of the standard model particles. 
However, \textit{the only interactions among pointons permitted by the Dirac Lagrangian $\mathcal{L}$---specifically, the mass terms of its chiral decomposition---is pair creation and annihilation of pointons of opposite charge.}
We therefore must account for each standard model trivalent vertex using only pointon pair creation and annihilation.

To see how geom interactions may work, let us consider the simplest interaction vertex: an electron-photon vertex.
In our composite model, electrons and positrons are pointons with future-directed and past-directed timelike orientations respectively, and a photon is a bound state of the two,
\begin{equation*}
e_a := [a0,00], \ \ \ \ \ \bar{e}_b := [0b,00], \ \ \ \ \ \gamma_a := [ab,00],
\end{equation*}
with subscripts $a,b \in \{ \uparrow, \downarrow \}$, $a \not = b$, denoting spin.
The timelike orientations arise from Hodge duality and not from an actual directed flow of time, since time is stationary along pointon worldlines; see Section \ref{charge subsection}. 
Furthermore, to obtain a classical description of the mass shell condition for pointons which does not violate relativity, the mass $m$ of a timelike pointon is \textit{not} defined as its rest energy $E_0$ (though $m = E_0$ holds precisely when it is on shell).
Rather, we define a pointon's mass to be the inverse radius $m := \hbar/(cr)$ of its `spinor particle' determined by its spin vector in its rest frame; see Section \ref{pointon spinor section} and \cite[Section 4]{B1}.
A pointon with a null worldline also has a spin vector, but, in contrast, cannot have a spinor particle, and thus has no mass. 
It therefore follows that although electrons and positrons are massive, photons need not be.

Using this composite model, let us consider two photons emitted from an electron in the Feynman-like diagram given in Figure \ref{electron picture}.
Observe that there are two electron-photon vertices in the diagram, and at each vertex two pointons of opposite charge are created.
Regardless of how far apart the two photons appear in any spatial hypersurface, they share a common pointon worldline (drawn in red), and thus are joined by a single `$1$-dimensional point' of spacetime. 
Consequently, if the photons each encounter a polarizer, then the two polarizers are effectively `touching' in spacetime, though not in any spatial hypersurface.
In this diagram, then, we obtain an approximate spacetime model of entanglement that is \textit{local in spacetime and nonlocal in space.} 

\begin{table} 
\caption{A composite model of the standard model particles with precisely one new massive spin-$2$ boson $x$, derived from the Dirac Lagrangian. 
Subscripts denote spin states, and $a,b \in \{\uparrow, \downarrow \}$.}
\label{particles}
\begin{center}
    \begin{tabular}{l||rrrr}
elec.\ charge & \multicolumn{4}{l}{all possible geoms $[\psi^-_1 \psi^+_1, \psi^-_2 \psi^+_2] = [\psi^-_2 \psi^+_2, \psi^-_1 \psi^+_1]$}\\
\hhline{=====}
$0$ & & $\gamma_{\uparrow} = [ \uparrow \downarrow, 00]$ & $Z_{\uparrow} = [ \uparrow \uparrow, 00]$ & $x_{ab} = [a 0, 0 b]$\\
(bosonic) & & $\gamma_{\downarrow} = [ \downarrow \uparrow, 00]$ & $Z_{\downarrow} = [ \downarrow \downarrow, 00]$ & $ x_0 = [ \downarrow \! {*}, {*} \! \downarrow]$ \\
& & $Z_0 = [ \uparrow \downarrow, \downarrow \uparrow]$ & $H = [\uparrow \uparrow, \downarrow \downarrow]$ & \\
\hdashline
$0$ & $\nu_e = [00, {*} \! \downarrow]$ & $\nu_{\mu} = [\downarrow \uparrow, {*} \! \downarrow]$ & $\nu_{\tau} = [\uparrow \uparrow, {*} \! \downarrow]$ & \\
(fermionic) & $\bar{\nu}_e = [00, \downarrow \! {*}]$ & $\bar{\nu}_{\mu} = [\uparrow \downarrow, \downarrow \! {*}]$ & $\bar{\nu}_{\tau} = [\uparrow \uparrow, \downarrow \! {*}]$ & \\
\hdashline
$-1$ &$e_{\uparrow} = [00, \uparrow \! 0]$ & $\mu_{\uparrow} = [\downarrow \uparrow, \uparrow \! 0]$ & $\tau_{\uparrow} = [ \downarrow \downarrow, \uparrow \! 0]$ & $\text{\footnotesize{$W^-_{\uparrow}$}} = [{*} \! \downarrow, \uparrow \! 0]$ \\
& $e_{\downarrow} = [00, \downarrow \! 0]$ & $\mu_{\downarrow} = [ \uparrow \downarrow, \downarrow \! 0]$ & $\tau_{\downarrow} = [\uparrow \uparrow, \downarrow \! 0]$ & $\text{\footnotesize{$W_{\downarrow}^-$}} = [{*} \! \downarrow, \downarrow \! 0]$ \\
 & & & & $\text{\footnotesize{$W_{0}^-$}} = [\downarrow \! {*}, \uparrow \! 0]$ \\
\hdashline
$+1$ & $\bar{e}_{\downarrow} = [00, 0 \! \uparrow ]$ & $\bar{\mu}_{\downarrow} = [\uparrow \downarrow, 0 \! \uparrow ]$ & $\bar{\tau}_{\downarrow} = [\downarrow \downarrow, 0 \! \uparrow]$ & $\text{\footnotesize{$W_{\downarrow}^+$}} = [\downarrow \! {*}, 0 \! \uparrow]$ \\
& $\bar{e}_{\uparrow} = [00, 0 \! \downarrow ]$ & $\bar{\mu}_{\uparrow} = [\downarrow \uparrow, 0 \! \downarrow]$ & $\bar{\tau}_{\uparrow} = [\uparrow \uparrow, 0 \! \downarrow]$ &
$\text{\footnotesize{$W_{\uparrow}^+$}} = [\downarrow \! {*}, 0 \! \downarrow]$\\
 & & & & $\text{\footnotesize{$W_{0}^+$}} = [{*} \! \downarrow, 0 \! \uparrow]$ \\
\hline
$\tfrac{-1}{3} + 1 = \tfrac{2}{3}$ & $u_{\uparrow} = \textcolor{red}{\pmb{(}} 00, {*} \! \downarrow]$ & $c_{\uparrow} = \textcolor{red}{\pmb{(}} \! \! \downarrow \uparrow, {*} \! \downarrow]$ 
& $t_{\uparrow} = \textcolor{red}{\pmb{(}} \! \! \uparrow \uparrow, {*} \! \downarrow]$ & \\
& $u_{\downarrow} = \textcolor{red}{\pmb{(}} 00, {*} \! \uparrow]$ & $c_{\downarrow} = \textcolor{red}{\pmb{(}} \! \! \uparrow \downarrow, {*} \! \uparrow]$ 
& $t_{\downarrow} = \textcolor{red}{\pmb{(}} \! \! \downarrow \downarrow, {*} \! \uparrow]$ &
\\
\hdashline
$\tfrac{1}{3} - 1 = \tfrac{-2}{3}$ & $\bar{u}_{\downarrow} = [00, \downarrow \! {*} \textcolor{red}{\pmb{)}}$ & $\bar{c}_{\downarrow} = [\uparrow \downarrow, \downarrow \! {*} \textcolor{red}{\pmb{)}}$ & $\bar{t}_{\downarrow} = [\uparrow \uparrow, \downarrow \! {*} \textcolor{red}{\pmb{)}}$ &
\\
 & $\bar{u}_{\uparrow} = [00, \uparrow \! {*} \textcolor{red}{\pmb{)}}$ & $\bar{c}_{\uparrow} = [\downarrow \uparrow, \uparrow \! {*} \textcolor{red}{\pmb{)}}$ & $\bar{t}_{\uparrow} = [\downarrow \downarrow, \uparrow \! {*} \textcolor{red}{\pmb{)}}$ &
\\
\hdashline
$\tfrac{-1}{3}$ & $d_{\uparrow} = \textcolor{red}{\pmb{(}} 00, \uparrow \! 0 ]$ & $s_{\uparrow} = \textcolor{red}{\pmb{(}} \! \! \downarrow \uparrow, \uparrow \! 0 ]$ & $b_{\uparrow} = \textcolor{red}{\pmb{(}} \! \! \downarrow \downarrow, \uparrow \! 0 ]$ & \\
&  $d_{\downarrow} = \textcolor{red}{\pmb{(}} 00, \downarrow \! 0 ]$ & $s_{\downarrow} = \textcolor{red}{\pmb{(}} \! \! \uparrow \downarrow, \downarrow \! 0 ]$ & $b_{\downarrow} = \textcolor{red}{\pmb{(}} \! \! \uparrow \uparrow, \downarrow \! 0 ]$ & \\
\hdashline
$\tfrac{1}{3}$ & $\bar{d}_{\downarrow} = [00, 0 \! \uparrow \! \! \textcolor{red}{\pmb{)}}$ & $\bar{s}_{\downarrow} = [\downarrow \uparrow, 0 \! \uparrow \! \! \textcolor{red}{\pmb{)}}$ & $\bar{b}_{\downarrow} = [\downarrow \downarrow, 0 \! \uparrow \! \! \textcolor{red}{\pmb{)}}$ &\\
& $\bar{d}_{\uparrow} = [00, 0 \! \downarrow \! \! \textcolor{red}{\pmb{)}}$ & $\bar{s}_{\uparrow} = [\uparrow \downarrow, 0 \! \downarrow \! \! \textcolor{red}{\pmb{)}}$ & $\bar{b}_{\uparrow} = [\uparrow \uparrow, 0 \! \downarrow \! \! \textcolor{red}{\pmb{)}}$ &\\
    \end{tabular}
  \end{center}
\end{table}

We consider the following combinatorial rules for geom interactions, with the aim of reproducing the standard model vertices. 
Orbitals and their fusions, introduced in \cite{B1}, are described in Sections \ref{pointon spinor section} and \ref{all orbital fusions}.
We call the pairs $(a_-,b_+)$ and $(b_-,a_+)$ of a geom ${[a_-a_+,b_-b_+]}$ its \textit{cross-orbitals}. 

\begin{Definition*} (Geom fusions.) \rm{
Two geoms $\delta_1 = {[ \psi_{11}, \psi_{12} ]}$ and $\delta_2 = {[ \psi_{21}, \psi_{22} ]}$ may \textit{fuse} to form a geom $\delta_3$, written ${\delta_1 \! \oslash \!  \delta_2} \to \delta_3$, if $\delta_3$ results from either
\begin{itemize}
 \item[(i)] two orbital fusions ${\psi_{1i} \! \oslash \!  \psi_{2i}}$, $i \in \{ 1,2 \}$, at least one of which is of the form
\begin{equation} \label{charged fusion1}
\text{\footnotesize{$
{\arraycolsep=2.9pt 
\begin{array}{c}
\xymatrix @C=-.3pc @R-2pc{
[ a & c \ar@{}[ld]^(.11){}="e"^(.95){}="f" \ar@{-} "e";"f" ]\\
[ c & b ] \\
\ar@{}[rr]^(-.45){}="a"^(1.0){}="b" \ar@{-} "a";"b" &&&&\\
[ a & b ] }
\end{array} }
$}}
\end{equation}
with $a,b \in \{ \uparrow, \downarrow \}$ and $c \in \{ 0, * \}$; or
 \item[(ii)] precisely one diagonal fusion ${\psi_{1i} \! \oslash \!  \psi_{2,i+1}}$, and $(*,0)$ is not a cross-orbital of $\delta_3$.
\end{itemize}
}\end{Definition*}

The diagonal line in (\ref{charged fusion1}) indicates that the two pointons labeled $c$ annihilate each other.
Moreover, (\ref{charged fusion1}) is a natural condition in the sense that it is a bit like sticking Lego pieces together.
The $(*,0)$ condition in (ii), however, warrants justification.
Our main theorem is the following.

\begin{Theorem*}
\ 
\begin{enumerate}
 \item There is a fusion for each standard model trivalent vertex, given in Table \ref{all interactions}, though with new constraints on the spin/polarization states at each vertex.
 \item There are precisely two fusions that do not correspond to standard model vertices:
\begin{equation} \label{new vertices}
Z_0 \! \oslash \!  Z_0 \to Z_0, \ \ \ \ \gamma_{\uparrow} \! \oslash \!  \gamma_{\downarrow} \to Z_0.
\end{equation}
\end{enumerate}
\end{Theorem*}

\begin{proof} 
Shown in Tables \ref{ffZ}, \ref{WWgamma}, \ref{parity}, and Theorem \ref{parity violation}.
\end{proof}

\begin{table} 
\caption{All standard model geom fusions. Here, $f$ is a charged fermion,  $\ell$ is a charged lepton, $d$ is a down-type quark, and $u$ is an up-type quark.
Subscripts $a,b,c \in \{ \uparrow, \downarrow \}$, $a \not = b$, denote spin states.}
\label{all interactions}
\begin{center}
$\begin{array}{rcrcrcr}
\hline
f_a \! \oslash \! \bar{f}_a \to \gamma_a & \ \ &
f_a \! \oslash \! \bar{f}_a \to Z_0 & \ \ &
f_a \! \oslash \! \bar{f}_b \to Z_a & \ \ &
\ell_{\uparrow} \! \oslash \! \bar{\nu}_{\ell} \to W^-_0
\\
 &&
\ell_a \! \oslash \! \bar{\ell}_b \to Z_0 &&
\nu_{\ell} \! \oslash \! \bar{\nu}_{\ell} \to Z_{\downarrow} &&
\bar{\ell}_{\downarrow} \! \oslash \! \nu_{\ell} \to W^+_0
\\
&&
d_a \! \oslash \! \bar{d}_b \to Z_0 &&
&&
d_{\uparrow} \! \oslash \! \bar{u}_{\uparrow} \to W^-_0
\\
W^-_a \! \oslash \! W^+_a \to \gamma_a &&
W^-_0 \! \oslash \! W^+_0 \to Z_0 &&
W^-_a \! \oslash \! W^+_b \to Z_a &&
\bar{d}_{\downarrow} \! \oslash \! u_{\downarrow} \to W^+_0
\\
\hline
\ell_a \! \oslash \! \bar{\ell}_c \to H &&
d_a \! \oslash \! \bar{d}_c \to H &&
W^-_a \! \oslash \! W^+_b \to H &&
Z_a \! \oslash \! Z_b \to H
\\
\nu_{\ell} \! \oslash \! \bar{\nu}_{\ell} \to H &&
u_a \! \oslash \! \bar{u}_a \to H &&
W^-_0 \! \oslash \! W^+_0 \to H &&
H \! \oslash \! H \to H
\\
\hline
\end{array}$
\end{center}
\end{table}

In Section \ref{differences}, we use the new geom fusions (\ref{new vertices}) to model the $4$-valent boson vertices and massive three-gluon vertices that occur in the standard model.

\textbf{Notation:} Tensors labeled with upper and lower indices $a,b, \ldots$ represent covector and vector slots respectively in Penrose's abstract index notation (so $v^a \in V$ and $v_a \in V^*$), and tensors labeled with indices $\mu, \nu, \ldots$ denote components with respect to a coordinate basis.
Given a curve $\beta: I \to \tilde{M}$, we often also denote its image $\beta(I)$ by $\beta$.
Natural units $\hbar = c = G = 1$ and the signature $(+,-,-,-)$ are used throughout.

\section{Preliminaries} \label{Preliminaries}
 
In this section we briefly review internal spacetime geometry and its resulting composite model of the standard model particles, introduced in \cite{B1,B6}.

\subsection{Internal spacetime geometry: charge and spin from orientation} \label{charge subsection}

Let ${(\tilde{M}, g = g_{ab})}$ be a $4$-dimensional orientable Lorentzian manifold, and denote its tangent spaces by $\tilde{M}_p$ for $p \in \tilde{M}$.
If $v$ is a unit timelike resp.\ null vector, then a minimal orthogonal projection of $v$ is
\begin{equation} \label{both projections}
[v]_{ab} := g_{ab} - v_av_b \ \ \ \ \text{ resp.\ } \ \ \ \ [v]_{ab} := g_{ab} + v_av'_b + v_bv'_a,
\end{equation}
where $v'$ is a null vector satisfying $v^a v'_a = -1$ \cite[Lemma 2.3]{B1}.

Internal spacetime is constructed by replacing the worldlines of dust particles with single `1-dimensional points': 

\begin{Definition} \rm{\cite[2.1, 2.2, 2.3]{B6}
Consider a countable set of dust particles on $\tilde{M}$ with worldlines $\beta_i \subset \tilde{M}$.
We call the set 
\begin{equation*}
M := (\tilde{M} \setminus (\cup_i \beta_i)) \cup (\cup_j \{\beta_j \}),
\end{equation*}
where each $\beta_j$ is a single point of $M$, 
an \textit{internal spacetime}, or simply \textit{spacetime}. 
We call $\tilde{M}$ the \textit{external spacetime}, $g$ an \textit{external metric}, and the dust particles \textit{pointons}.

Fix $p \in \tilde{M}$.
Let $v_1, \ldots, v_n \in \tilde{M}_p$ be the tangent vectors to the pointon worldlines $\beta_1, \ldots, \beta_n$ at $p$.
We define the \textit{internal metric} to be the degenerate symmetric rank-$2$ tensor given by the composition of projections (\ref{both projections}),\footnote{The external metric $g_{ab}$ is used to identify the tangent and cotangent spaces $\tilde{M}_p$ and $\tilde{M}_p^*$, and thus to raise and lower indices.} 
\begin{equation*} \label{projection}
h = h_p = \tensor{h}{^a_b} := \tensor{[v_1]}{^a_c}\tensor{[v_2]}{^c_d} \cdots \tensor{[v_n]}{^e_b}: \tilde{M}^*_p \otimes \tilde{M}_p \to \mathbb{R}.
\end{equation*}
The \textit{(internal) tangent space} at a point $p \in \tilde{M}$ is the image of $h$ at $p$,
\begin{equation*}
M_p := \im h = \im (\tensor{h}{^a_b}) = \{ v^a \in \tilde{M}_p \, | \, \tensor{h}{^a_b}\tensor{v}{^b} = \tensor{v}{^a} \} \subseteq \tilde{M}_p.
\end{equation*}
We call a vector $v \in \tilde{M}_p$ \textit{internal} if $v$ is in $M_p$, and \textit{external} otherwise. 
}\end{Definition}

The metric on $M$ is defined so that the $\beta_i$ are \textit{not} contracted to $0$-dimensional points, in contrast to geometric and topological quotients.
Degenerate spacetime metrics also appear in the contexts of loop quantum gravity \cite{LW}, and as a possible means of regularizing the big bang singularity \cite{K1, K2, K3, KW, Ba}.
The geometric framework considered here was recently introduced to study certain (`nonnoetherian') rings in algebraic geometry \cite{B3,B4,B5}.

Let $\beta \subset \tilde{M}$ the worldline of a pointon.
Since $\beta$ is a single point of spacetime $M$, its pointon does not have a $4$-velocity $v$ in $M$.
We thus must replace $v$ with a new geometric object that is fully intrinsic to $M$.

A natural candidate is the Hodge dual to the volume form $\vol (\ker h)$ of the kernel of $h$, since it is wedge product of $1$-forms that live in the internal tangent space $M_{\beta(t)}^*$.
However, Hodge duals are \textit{pseudo}-forms, and thus depend on a choice of orientation $o_{\tilde{M}} \in \{ \pm 1 \}$ of $\tilde{M}_{\beta(t)}$.
Since our new geometric object cannot depend on $\tilde{M}$, we acquire a new $\mathbb{Z}_2$ parameter $o_{\ker h} \in \{ \pm 1\}$ that replaces the orientation of the vanishing subspace of $h$ and is independent of $\tilde{M}$.

\begin{Definition} \cite[3.1]{B6} \rm{
We define the \textit{internal $4$-velocity} of a pointon at $p = \beta(t) \in \tilde{M}$ with $4$-velocity $v \in \ker h \subset \tilde{M}_p$ to be the pseudo-form
\begin{equation*} \label{internal form}
\breve{v}_{a \cdots b} := o_{\operatorname{ker}h} \star \! \vol (\ker h) \in {\bigwedge} \! ^{\dim M_p} \, M_p^*,
\end{equation*}
where $o_{\operatorname{ker}h} \in \{ \pm 1 \}$ is a free parameter independent of any orientation of $\tilde{M}_p$.
Observe that the rank of $\breve{v}$ changes along $\beta \subset \tilde{M}$ whenever the dimension of $M_{\beta(t)}$ changes.
}\end{Definition}

\begin{table}
\label{charge and spin}
\caption{Charge and spin from vanishing subspaces of spacetime tangent spaces.
Here we assume that the nonphysical choice of orientation $o_{\tilde{M}} = o_{0123} = o_0o_1o_2o_3 \in \{ \pm 1 \}$ of $\tilde{M}_{\beta(t)}$ is $1$.}
\begin{center}
\begin{tabular}{|l|l|l|}
\hline
vanishing subspace & free orientation & identification\\
\hline \hline
$e_0$ & $o_0 = o_{123}$ & electric charge $e^{o_0}$\\
\hdashline
$e_0 \wedge e_2 \wedge e_3$ & $o_1 = o_{023}$ & color charge $r^{o_1}$\\
$e_0 \wedge e_3 \wedge e_1$ & $o_2 = o_{013}$ & color charge $g^{o_2}$\\
$e_0 \wedge e_1 \wedge e_2$ & $o_3 = o_{012}$ & color charge $b^{o_3}$\\
\hdashline
$e_2 \wedge e_3$ & $o_{23} = o_{01}$ & spin in the direction $o_{01}e_1$\\
$e_3 \wedge e_1$ & $o_{13} = o_{02}$ & spin in the direction $o_{02}e_2$\\
$e_1 \wedge e_2$ & $o_{12} = o_{03}$ & spin in the direction $o_{03}e_3$\\
\hline
\end{tabular}
\end{center}
\end{table}

Suppose a pointon has timelike $4$-velocity $v$ and worldline $\beta$ in $\tilde{M}$, and let $e_0, \ldots, e_3$ be a tetrad parallel transported along $\beta$ for which $e_0 = v$.
Fix a point $p = \beta(t) \in \tilde{M}$.

If the pointon is isolated at $p$, then $\breve{v}_{abc} = o_0 e^1 \wedge e^2 \wedge e^3$, where $o_0 \in \{ \pm 1 \}$ is a free choice of time orientation that we identify with the electric charge of the pointon.

If instead $\beta$ transversely intersects another pointon worldline at $p$, then the dimension of the tangent spaces $M_{\beta(t)}$ along $\beta$ drops at $p$.
In the case $\dim M_p = 1$, the internal $4$-velocity is an actual spatial vector, say $\breve{v}_a = o_0o_{12}e^3$, with $o_{12} \in \{ \pm 1\}$ a free choice of orientation of the plane spanned by $e_1$ and $e_2$ in $\tilde{M}_p$.
We call this vector the \textit{spin vector} of the pointon, and denote its parallel transport along $\beta$ by $s$.
We say the pointon has spin up resp.\ down in the $e_3$ direction if $o_{12} = 1$ resp.\ $o_{12} = -1$. 
Spin wavefunction collapse occurs when $s$ is projected under $h$ onto another $1$-dimensional tangent space $M_{\beta(t')}$, for some $t' > t$.
This degenerate spacetime geometry yields the Kochen-Specker model of spin \cite{KS}, and thus reproduces the Born rule for spin wavefunction collapse \cite[Section 4]{B6}.

In general, spin arises from the orientation of a vanishing $2$-dimensional subspace, and charge arises from the orientation of a vanishing $1$- or $3$-dimensional subspace.
These identifications are given in Table \ref{charge and spin}.

\subsection{The spinor of a pointon} \label{pointon spinor section}

Consider a pointon with worldline $\beta \subset \tilde{M}$, timelike $4$-velocity $v \in \tilde{M}_{\beta(t)}$, and spin vector $s \in M_{\beta(t)}$.
Recall that pointons are dust particles, and thus possess energy. 
We would like to identify the pointon's rest energy $\omega > 0$, or equivalently $4$-momentum $k = \omega v$, with something geometric. 

Observe that the \textit{internal $4$-momentum} of the pointon is
\begin{equation*}
\breve{k}_{abc} := o_0 \star \! k_d = o_0 \star \! \omega v_d = \omega \breve{v}_{abc}.
\end{equation*}
Denote by $\hat{\star}$ the Hodge dual in the internal tangent space $M_{\beta(t)}$.
Then the internal Hodge dual of the contraction $s^a \breve{k}_{abc}$ is the vector
\begin{equation*}
\hat{\star} s^a \breve{k}_{abc} = \omega s,
\end{equation*}
and therefore may be interpreted as an angular frequency vector in the rest frame of the pointon \cite[Lemma 3.1]{B1}.
A simplest proposal would be that $\omega s$ is the orbital angular velocity vector of a particle circling the pointon.
We call such a particle with worldline $\alpha \subset \tilde{M}$ the \textit{spinor particle} of the pointon.

Recall the gamma matrix map in the chiral basis, 
\begin{center}
$\begin{array}{rcl}
\gamma: \mathbb{R}^{1,3} 
& \longrightarrow & M_4(\mathbb{C})\\
v & \mapsto & \left[ \begin{smallmatrix} & v_* \\ v^* & \end{smallmatrix} \right]_{\mathscr{C}} = \gamma(v^{\mu}e_{\mu})  = v^{\mu} \gamma(e_{\mu}) = v^{\mu} \gamma_{\mu} = \slashed v,
\end{array}$
\end{center} 
where
\begin{equation*}
v_* := v_0 \bm{1} + v_j \sigma^j = \left[ \begin{smallmatrix} v_0 + v_3 & v_1 - i v_2 \\ v_1 + i v_2 & v_0 - v_3 \end{smallmatrix} \right]
\ \ \ \text{ and } \ \ \ 
v^* := v_0 \bm{1} - v_j \sigma^j = \left[ \begin{smallmatrix} v_0 - v_3 & -v_1 + i v_2 \\ -v_1 - i v_2 & v_0 + v_3 \end{smallmatrix} \right].
\end{equation*}
We define the \textit{spinor} $\psi(\beta(t)) \in \mathbb{C}^4$ of a charged pointon to be one of the four columns of the gamma matrix of the internal metric $h$ of the $4$-velocity $\dot{\alpha}^{\mu} := \tfrac{d}{d\tau}\alpha^{\mu}$ of its spinor particle,
\begin{equation*}
\gamma(h(\dot{\alpha})) = \gamma^{\mu}h_{\mu \nu} \dot{\alpha}^{\nu}.
\end{equation*}

We say a set of spinors $\mathcal{O}$ is \textit{covariantly independent} if for any two spinors $\psi_1, \psi_2 \in \mathcal{O}$, we have $\psi_1 = \psi_2$ whenever there is a $c \in \mathbb{C}$ and $\Lambda \in L_0$ such that $\psi_1 = c \Lambda \psi_2$ \cite[Definition 3.4]{B1}.
The spinors in the chiral basis $\mathscr{C}$,
\begin{equation*} \label{O5}
[ \uparrow \! 0] := \left[ \begin{smallmatrix} e^{i \omega t} \\ 0 \\ 0 \\ 0 \end{smallmatrix} \right]_{\mathscr{C}}, \ \ \ 
[ \downarrow \! 0] := \left[ \begin{smallmatrix} e^{-i \omega t} \\ 0 \\ 0 \\ 0 \end{smallmatrix} \right]_{\mathscr{C}}, \ \ \ 
[ 0 \! \uparrow ] := \left[ \begin{smallmatrix} 0 \\ 0 \\ -e^{-i \omega t} \\ 0 \end{smallmatrix} \right]_{\mathscr{C}}, \ \ \ 
[ 0 \! \downarrow ] := \left[ \begin{smallmatrix} 0 \\ 0 \\ -e^{i \omega t} \\ 0 \end{smallmatrix} \right]_{\mathscr{C}},
\end{equation*}
form a maximal covariantly independent set of charged pointon spinors, which we denote $\mathcal{O}_5$ \cite[Proposition 3.5]{B1}.
Here, each spinor is a $\gamma^5$ eigenspinor, written in the rest frame of its pointon $\beta \subset \tilde{M}$ with spin in the $e_3$ direction.
Furthermore, these spinors are solutions to the Klein-Gordon equation, 
\begin{equation*}
(\partial^2 + \omega^2)\psi = (-i \slashed \partial - \omega)(i \slashed \partial - \omega) \psi = 0.
\end{equation*}
We thus obtain the Klein-Gordon equation from the internal metric $h$ together with the gamma representation of $\tilde{M}_{\beta(t)}$, \textit{without assuming canonical quantization}.

\begin{figure} \label{spin imped}
\begin{equation*}
\begin{array}{ccccccccccc}
\begin{tikzpicture}[scale=3,cap=round,>=latex]
  \draw (0cm,0cm) circle[radius=.2cm];
\fill[radius=.03cm] (0cm,.2cm) circle[];
   \draw[blue, line width=.7mm, opacity=.5,->] (0cm,0cm)+(83:.2cm) arc[start angle=83, end angle=30, radius=.2cm];
 \draw (0cm,-.225cm) node[anchor=north] {${[ \uparrow \! 0]}$};
\end{tikzpicture}
&&
\begin{tikzpicture}[scale=3,cap=round,>=latex]
  \draw (0cm,0cm) circle[radius=.2cm];
\fill[radius=.03cm] (0cm,.2cm) circle[];
   \draw[blue, line width=.7mm, opacity=.5,->] (0cm,0cm)+(97:.2cm) arc[start angle=97, end angle=150, radius=.2cm];
 \draw (0cm,-.225cm) node[anchor=north] {${[ \downarrow \! 0]}$};
\end{tikzpicture}
&&
\begin{tikzpicture}[scale=3,cap=round,>=latex]
  \draw (0cm,0cm) circle[radius=.2cm];
  \draw (0cm,.2cm) circle[radius=.03cm];
   \draw[blue, line width=.7mm, opacity=.5,->] (0cm,0cm)+(97:.2cm) arc[start angle=97, end angle=150, radius=.2cm];
 \draw (0cm,-.225cm) node[anchor=north] {${[0 \! \uparrow]}$};
\end{tikzpicture}
&&
\begin{tikzpicture}[scale=3,cap=round,>=latex]
  \draw (0cm,0cm) circle[radius=.2cm];
 \draw (0cm,.2cm) circle[radius=.03cm];
   \draw[blue, line width=.7mm, opacity=.5,->] (0cm,0cm)+(83:.2cm) arc[start angle=83, end angle=30, radius=.2cm];
 \draw (0cm,-.225cm) node[anchor=north] {${[0 \! \downarrow]}$};
\end{tikzpicture}
& &
\begin{tikzpicture}[scale=3,cap=round,>=latex]
  \draw (0cm,0cm) circle[radius=.2cm];
  \draw (0cm,-.2cm) circle[radius=.03cm];
\fill[radius=.03cm] (0cm,.2cm) circle[];
   \draw[blue, line width=.7mm, opacity=.5,->] (0cm,0cm)+(83:.2cm) arc[start angle=83, end angle=30, radius=.2cm];
   \draw[blue, line width=.7mm, opacity=.5,->] (0cm,0cm)+(-83:.2cm) arc[start angle=-83, end angle=-30, radius=.2cm];
 \draw (0cm,-.225cm) node[anchor=north] {$[ \uparrow \uparrow]$};
\end{tikzpicture}
& 
\raisebox{3.1 em}{ \ }
&
\begin{tikzpicture}[scale=3,cap=round,>=latex]
  \draw (0cm,0cm) circle[radius=.2cm];
  \draw (0cm,-.2cm) circle[radius=.03cm];
\fill[radius=.03cm] (0cm,.2cm) circle[];
   \draw[blue, line width=.7mm, opacity=.5,->] (0cm,0cm)+(97:.2cm) arc[start angle=97, end angle=150, radius=.2cm];
   \draw[blue, line width=.7mm, opacity=.5,->] (0cm,0cm)+(-97:.2cm) arc[start angle=-97, end angle=-150, radius=.2cm];
 \draw (0cm,-.225cm) node[anchor=north] {$[ \downarrow \downarrow]$};
\end{tikzpicture}
\\
\begin{tikzpicture}[scale=3,cap=round,>=latex]
  \draw (0cm,0cm) circle[radius=.2cm];
\fill[pink, radius=.03cm] (0cm,.2cm) circle[];
\draw (0cm,.2cm) circle[radius=.03cm];
   \draw[blue, line width=.7mm, opacity=.5,->] (0cm,0cm)+(97:.2cm) arc[start angle=97, end angle=150, radius=.2cm];
 \draw (0cm,-.225cm) node[anchor=north] {$[ \ \uparrow \ ]$};
\end{tikzpicture}
& 
\raisebox{3.1 em}{$=$}
&
\begin{tikzpicture}[scale=3,cap=round,>=latex]
  \draw (0cm,0cm) circle[radius=.2cm];
  \draw (0cm, 0cm) circle[radius=.03cm];
\fill[radius=.03cm] (0cm,.2cm) circle[];
   \draw[blue, line width=.7mm, opacity=.5,->] (0cm,0cm)+(97:.2cm) arc[start angle=97, end angle=150, radius=.2cm];
 \draw (0cm,-.225cm) node[anchor=north] {$[ \downarrow \! {*}]$};
\end{tikzpicture}
&
\raisebox{3.1 em}{ \ }
&
\begin{tikzpicture}[scale=3,cap=round,>=latex]
  \draw (0cm,0cm) circle[radius=.2cm];
\fill[pink, radius=.03cm] (0cm,.2cm) circle[];
\draw (0cm,.2cm) circle[radius=.03cm];
  \draw[blue, line width=.7mm, opacity=.5,->] (0cm,0cm)+(83:.2cm) arc[start angle=83, end angle=30, radius=.2cm];
 \draw (0cm,-.225cm) node[anchor=north] {$[ \ \downarrow \  ]$};
\end{tikzpicture}
& 
\raisebox{3.1 em}{$=$}
&
\begin{tikzpicture}[scale=3,cap=round,>=latex]
  \draw (0cm,0cm) circle[radius=.2cm];
\fill[radius=.03cm] (0cm,0cm) circle[];
 \draw (0cm,.2cm) circle[radius=.03cm];
   \draw[blue, line width=.7mm, opacity=.5,->] (0cm,0cm)+(83:.2cm) arc[start angle=83, end angle=30, radius=.2cm];
 \draw (0cm,-.225cm) node[anchor=north] {$[{*} \! \downarrow]$};
\end{tikzpicture}
&
\raisebox{3.1 em}{ \ }
&
\begin{tikzpicture}[scale=3,cap=round,>=latex]
  \draw (0cm,0cm) circle[radius=.2cm];
\draw (0cm,.21cm) circle[radius=.03cm];
\fill[radius=.03cm] (0cm,.19cm) circle[];
 \draw (0cm,-.225cm) node[anchor=north] {$[ \uparrow \downarrow]$};
   \draw[blue, line width=.7mm, opacity=.5,->] (0cm,0cm)+(83:.2cm) arc[start angle=83, end angle=30, radius=.2cm];
\end{tikzpicture}
 & &
\begin{tikzpicture}[scale=3,cap=round,>=latex]
  \draw (0cm,0cm) circle[radius=.2cm];
\draw (0cm,.21cm) circle[radius=.03cm];
\fill[radius=.03cm] (0cm,.19cm) circle[];
 \draw (0cm,-.225cm) node[anchor=north] {$[ \downarrow \uparrow]$};
   \draw[blue, line width=.7mm, opacity=.5,->] (0cm,0cm)+(97:.2cm) arc[start angle=97, end angle=150, radius=.2cm];
\end{tikzpicture}
\end{array}
\end{equation*}
\caption{All geom orbitals, depicted as spinor particles.}
\end{figure}

Denote by $r$ the radius of the spinor particle's trajectory, and by $u = \omega r$ its tangential speed in the rest frame of the pointon.
We define the \textit{mass} $m$ of the pointon to be the special frequency $\omega_*$ for which $u$ is lightlike, $u = 1$; whence $m = \omega_* = r^{-1} = \hbar/(cr)$. 
Putting $E_0 = \omega$, $u = r \omega$, and $m = r^{-1}$ together we obtain
\begin{equation} \label{ngutt}
E_0 = \omega = ur^{-1} = mu = mcu.
\end{equation}
If $u = c = 1$, then we recover $E_0 = mc^2 = m$.
Furthermore, (\ref{ngutt}) implies that the $4$-momentum $k^a$ of any timelike pointon satisfies
\begin{equation} \label{k^2}
k^2 = k^a k_a = E_0^2 = \omega^2 = m^2u^2.
\end{equation}
Consequently, a pointon is on shell, $k^2 = m^2$, if and only if $u = 1$, whereas the relativistic constraint $k^2 = E_0^2$ holds even if the pointon is off shell.
\textit{We therefore obtain a model of off-shell particles for which relativity is never violated.} 

The relation (\ref{k^2}) modifies the Dirac Lagrangian $\mathcal{L}_{\mathscr{D}} = \bar{\psi}(i \slashed \partial - m) \psi$, $\bar{\psi} := \psi^{\dagger}\gamma^0$, by replacing $m$ with $mu = r^{-1}u = \omega$:
\begin{equation*}
\mathcal{L} = \bar{\psi}(i \slashed \partial - mu) \psi = \bar{\psi}(i \slashed \partial - \omega) \psi.
\end{equation*}
Thus, since $u$ is a free parameter, off-shell spinors are solutions to the \textit{classical equations of motion} of $\mathcal{L}$: $(i \slashed \partial - mu) \psi = 0$.
From this we define the \textit{pointon Dirac Lagrangian} to be
\begin{equation*}
\mathcal{L} = \bar{\psi}(i \slashed \partial - \hat{\omega}) \psi,
\end{equation*}
where $\hat{\omega}$ is a $\mathbb{C}$-linear operator that acts on a pointon spinor $\psi$ with rest energy $\omega > 0$ and spin orientation $o_P \in \{ \pm 1 \}$ by $\hat{\omega}\psi := o_P\omega \psi$ \cite[Section 5]{B1}.

It was shown in \cite[Proposition 2.2]{B1} that the two direct summands of the Lie algebra decomposition 
\begin{equation*}
\mathfrak{so}(1,3)_{\mathbb{C}} \cong \mathfrak{su}(2)_{\mathbb{C}} \oplus \mathfrak{su}(2)_{\mathbb{C}}
\end{equation*}
correspond to time orientations $o_0 \in \{ \pm 1 \}$, and thus to positive and negative electric charges. 
Therefore, \textit{chirality is electric charge for spinors on an internal spacetime}.
Thus, if $\psi \in \mathbb{C}^4$ is an eigenspinor of $\gamma^5$ with eigenvalue $\pm 1$, then $\psi$ has electric charge $\pm 1$; otherwise $\psi$ is neutral.
Consequently the projections
\begin{equation*}
\psi^- := \tfrac 12(1 - \gamma^5) \psi = \left[ \begin{smallmatrix} * \\ * \\ 0 \\ 0 \end{smallmatrix} \right]_{\mathscr{C}} \ \ \ \ \text{ and } \ \ \ \ \psi^+ := \tfrac 12 (1 + \gamma^5) \psi = \left[ \begin{smallmatrix} 0 \\ 0 \\ * \\ * \end{smallmatrix} \right]_{\mathscr{C}}
\end{equation*}
are spinors with charges $-1$ and $+1$, respectively.
In particular, the pointon spinors ${[a0]}$ and ${[0a]}$, $a \in \{ \uparrow, \downarrow \}$, have negative and positive charges respectively.

This new identification of chirality significantly changes the meaning of the Dirac Lagrangian.
Indeed, the mass terms of its chiral decomposition,
\begin{equation*}
\mathcal{L} = \bar{\psi}(i \slashed \partial - \hat{\omega}) \psi = i \bar{\psi}^- \slashed \partial \psi^- - \hat{\omega} \bar{\psi}^- \psi^+ + (+ \leftrightarrow -),
\end{equation*}
now represent
\begin{itemize}
 \item[(\textsc{a})] couplings between pointons of opposite charge; and
 \item[(\textsc{b})] pair creation/annihilation of pointons of opposite charge, that is, two-valent interaction vertices.
\end{itemize}
Set $\delta : = (i \slashed \partial - \hat{\omega})$.

From (\textsc{a}), a spinor $\psi = \psi^- + \psi^+$ is a bound state of its charged summands $\psi^-$ and $\psi^+$ if and only if 
\begin{equation} \label{bound state}
\delta(\psi^- + \psi^+)(p) = \delta \psi(p) = 0
\end{equation}
for periodic $p \in \tilde{M}$.\footnote{
The coupling of pointons of opposite charge described by the Dirac equation is similar to the coupling of electric and magnetic fields described by Maxwell's equations.
Indeed, let $[\psi^-]$ (resp.\ $[\psi^+]$) denote the upper (resp.\ lower) two components of $\psi^-$ (resp.\ $\psi^+$).
Then the chiral Dirac equations in the pointon's rest frame, namely
\begin{equation*}
\tfrac 1c \partial_0 [\psi^+] = \pm \tfrac ir  [\psi^-] \ \ \ \ \text{ and } \ \ \ \ \tfrac 1c \partial_0 [\psi^-] = \pm \tfrac ir [\psi^+],
\end{equation*}
are markedly similar to the Faraday and Amp\'ere equations,
\begin{equation*}
\tfrac 1c \partial_0 \bm{B} = - \nabla \! \times \! \bm{E} \ \ \ \ \text{ and } \ \ \ \ \tfrac 1c \partial_0 \bm{E} = \nabla \! \times \! \bm{B},
\end{equation*}
since $[\psi^-]$ and $[\psi^+]$ \textit{specify the rotation, or phase, of a spinor particle}.}
The spinors in $\mathcal{O}_5$, in particular, are not bound states of pointons of opposite charge \cite[Proposition 3.5]{B1}, as expected.
In contrast to standard quantum theory, spinor solutions to the Dirac equation on an internal spacetime are bound states of two oppositely charged pointons.
Thus, since pointons are fermions, \textit{(periodic) Dirac solutions are bosons}.\footnote{A Cooper pair is an example of a bound state of two fermions that is bosonic.}

From (\textsc{b}), we say that two spinors $\psi_1$, $\psi_2$ may couple, or \textit{fuse} at $p \in \tilde{M}$, if whenever $\psi_1$, $\psi_2$ are charged resp.\ neutral we have
\begin{equation} \label{coupling constraints}
\delta(\psi_1^{\pm} + \psi_2^{\mp})(p) = 0 \ \ \ \text{ resp.\ } \ \ \ \delta(\psi_1 + \psi_2)(p) = 0
\end{equation}
and 
\begin{equation*}
\delta \psi_1 \equiv 0 \ \ \Longleftrightarrow \ \ \delta \psi_2 \equiv 0.
\end{equation*}
Two fusible spinors $\psi_1 = [ab]$ and $\psi_2 = [cd]$, with $a,b,c,d \in \{ 0, \uparrow, \downarrow \}$, may fuse into one of three spinors $[ad]$, $[cb]$, $[00]$:
\begin{equation*}
\text{\footnotesize{$
{\arraycolsep=2.9pt 
\begin{array}{ccc}
\xymatrix @C=-.3pc @R-2pc{
[ a & b \ar@{}[ld]^(.11){}="e"^(.95){}="f" \ar@{-} "e";"f" ]\\
[ c & d ] \\
\ar@{}[rr]^(-.45){}="a"^(1.0){}="b" \ar@{-} "a";"b" &&&&\\
[ a & d ] }
&
\xymatrix @C=-.3pc @R-2pc{
[ a \ar@{}[rd]^(.05){}="e"^(.9){}="f" \ar@{-} "e";"f" & b ]\\
[ c & d ] \\
\ar@{}[rr]^(-.45){}="a"^(1.0){}="b" \ar@{-} "a";"b" &&&&\\
[ c & b ] }
&
\xymatrix @C=-.3pc @R-2pc{
[ a \ar@{}[rd]^(.05){}="e"^(.9){}="f" \ar@{-} "e";"f" & b \ar@{}[ld]^(.11){}="e"^(.95){}="f" \ar@{-} "e";"f" ]\\
[ c & d ] \\
\ar@{}[rr]^(-.45){}="a"^(1.0){}="b" \ar@{-} "a";"b" &&&&\\
[ 0 & 0 ] }
\end{array} }
$}}
\end{equation*}
where the diagonal lines indicate pointon pair annihilation $\bar{\psi}^{-} \hat{\omega} \psi^{+}$.

The spinors $[\uparrow \downarrow ]$ and $[\downarrow \uparrow ]$ are $\gamma^0$ eigenspinors, and similar to $\mathcal{O}_5$, we consider a maximal covariantly independent set $\mathcal{O}_0$ of $\gamma^0$ eigenspinors,
\begin{equation*}
[ \downarrow \! *] := \left[ \begin{smallmatrix} e^{i \omega t} \\ 0 \\ 0 \\ 0 \end{smallmatrix} \right]_{\mathscr{D}}, \ \ \ 
 [ * \! \downarrow] := \left[ \begin{smallmatrix} e^{-i \omega t} \\ 0 \\ 0 \\ 0 \end{smallmatrix} \right]_{\mathscr{D}}, \ \ \ 
[ \downarrow \uparrow ] = \left[ \begin{smallmatrix} 0 \\ 0 \\ -e^{-i \omega t} \\ 0 \end{smallmatrix} \right]_{\mathscr{D}}, \ \ \ 
[ \uparrow \downarrow ] = \left[ \begin{smallmatrix} 0 \\ 0 \\ -e^{i \omega t} \\ 0 \end{smallmatrix} \right]_{\mathscr{D}},
\end{equation*}
where the subscript $\mathscr{D}$ denotes the Dirac basis \cite[Proposition 5.4]{B1}.\footnote{The change-of-basis matrices are $\tfrac{1}{\sqrt{2}} \! \left[ \begin{smallmatrix} \bm{1} & -\bm{1} \\ \bm{1} & \bm{1} \end{smallmatrix} \right] \psi_{\mathscr{D}} = \psi_{\mathscr{C}}$ and 
$\tfrac{1}{\sqrt{2}} \! \left[ \begin{smallmatrix} \bm{1} & \bm{1} \\ -\bm{1} & \bm{1} \end{smallmatrix} \right] \psi_{\mathscr{C}} = \psi_{\mathscr{D}}$.}
The spinors ${[\downarrow \! *]}$ and ${[* \! \downarrow]}$ are not bound states of two pointons of opposite charge by (\ref{bound state}), nor $\gamma^5$ eigenspinors, and thus are single neutral pointon spinors \cite[Lemma 5.5]{B1}.
Consequently, they admit only one fusion. 
Furthermore, since ${[\downarrow \! *]}$ and ${[* \! \downarrow]}$ have opposite angular momentum, their fusion must yield a neutral spinor ${[ab]}$ with zero angular momentum.
There are only two such choices, namely ${[\uparrow \uparrow]}$ and ${[\downarrow \downarrow ]}$, and so we choose, once and for all, the latter:
\begin{equation*}
\text{\footnotesize{$
{\arraycolsep=2.9pt 
\begin{array}{c}
\xymatrix @C=-.3pc @R-2pc{
[ \downarrow & {*} \ar@{}[ld]^(.11){}="e"^(.95){}="f" \ar@{-} "e";"f" ]\\
[ {*} & \downarrow ] \\
\ar@{}[rr]^(-.45){}="a"^(1.0){}="b" \ar@{-} "a";"b" &&&&\\
[ \downarrow & \downarrow ] }
\end{array} }
$}}
\end{equation*}

To summarize, the pointon spinors with negative (resp.\ positive; zero) charge are $[a0]$ (resp.\ $[0a]$; ${[\downarrow \! *]}$, ${[* \! \downarrow ]}$), with $a \in \{ \uparrow, \downarrow \}$. 
We call pointon spinors and spinors obtained from any number of their fusions \textit{orbitals}; the set of all orbitals is then $\mathcal{O}_0 \cup \mathcal{O}_5 \cup \{[\uparrow \uparrow], [\downarrow \downarrow ]\}$. 

\begin{Theorem} \label{allorbitals} \cite[Theorem 5.6]{B1}
The pairs of orbitals that may fuse are precisely
\begin{equation*} 
\{ [a0], [0b] \}, \ \ \ \ \{ [ab], [ab] \}, \ \ \ \ \{ [ab], [ba] \}, \ \ \ \ \{ [aa], [bb] \}, \ \ \ \ \{ [\downarrow \! *], [* \! \downarrow] \},
\end{equation*}
with $a,b \in \{ \uparrow, \downarrow \}$. 
\end{Theorem}

Two orbitals $\psi_1 = [ab], \psi_2 = [cd]$, $a,b,c,d \in \{  0, *, \uparrow, \downarrow \}$, may couple from the mass term $\bar{\psi}_1 \hat{\omega} \psi_2$ of $\mathcal{L} = \bar{\psi}_1 \delta \psi_2$ to form a bound state, which we denote $[ab,cd] = [cd,ab]$, whenever it does not violate the Pauli exclusion principle. 
We call these bound pairs \textit{geoms} for `geometric atom', and identify the pairs with the elementary particles of the standard model; see Table \ref{particles}.
Note that a geom consists of a most four pointons.
Furthermore, the electric charge of a geom is the sum of its constituent charges, and a geom is a fermion (resp.\ boson) if it contains an odd (resp.\ even) number of pointons.

\subsection{Color charge}

In \cite{B2} we introduce an internal spacetime geometric model of color charge for pointons.
In this model, color charge is a free spacelike orientation
\begin{equation*}
o_1 = o_{023} \sim r^{o_1}, \ \ \ \ \ o_2 = o_{013} \sim g^{o_2}, \ \ \ \ \ o_3 = o_{012} \sim b^{o_3},
\end{equation*}
that arises, via the Hodge dual pseudo-form of the pointon's $4$-velocity, from the vanishing of a $(2+1)$-dimensional hypersurface; see Table \ref{charge and spin}.
Gluons are modeled similar to flux tubes (or strings), and in particular are not geoms. 
Furthermore, we find that
\begin{itemize}
 \item[(\textsc{i})] only orbitals with a single pointon may have nonzero color charge; and
 \item[(\textsc{ii})] a geom can have at most one unit of color charge, just as with electric charge. 
\end{itemize}

The only orbitals that may possess color charge are therefore ${[a0]}$, ${[0a]}$, with $a \in \{ \uparrow, \downarrow \}$, and ${[\downarrow \! *]}$, ${[* \! \downarrow ]}$.
These colored orbitals are denoted
\begin{equation*}
\textcolor{red}{\pmb{(}} a0], \ \ \ \ [0 a \textcolor{red}{\pmb{)}}, \ \ \ \ [\downarrow \! * \textcolor{red}{\pmb{)}}, \ \ \ \ \textcolor{red}{\pmb{(}} * \! \downarrow ],
\end{equation*}
where a colored left (resp.\ right) rounded bracket indicates color charge $c^-_i$ (resp.\ $c^+_i$).
It also follows from our model that the rounded bracket must be adjacent to any $*$ component; indeed, without this placement our model would not reproduce the $udW$ vertices given in Table \ref{parity} below. 
Furthermore, colored $*$-orbitals are spin states of a \textit{charged} spin-$\tfrac 12$ particle, and there are thus four such states (positive/negative charge and spin up/down). 
We therefore must allow $\uparrow$ arrows in colored $*$-orbitals, yielding 
\begin{equation*} \label{four}
[\downarrow \! * \textcolor{red}{\pmb{)}}, \ \ \ \ \textcolor{red}{\pmb{(}} * \! \uparrow], \ \ \ \ 
\textcolor{red}{\pmb{(}} * \! \downarrow], \ \ \ \ [\uparrow \! * \textcolor{red}{\pmb{)}}.
\end{equation*}
	
A geom with total color charge $c^-_i$ (resp.\ $c^+_i$) is indicated by a colored left (resp.\ right) rounded bracket, just as for orbitals.
To obtain the electric charge of a geom with color, substitute $\pm \tfrac 13$ for each colored geom component with electric charge $\pm 1$, $c_i^{o_i} \mapsto \tfrac 13 e^{o_i}$, and then sum the electric charges of each component.

We will discuss our replacement of the three-gluon vertex in Section \ref{differences}.

\subsection{All orbital fusions} \label{all orbital fusions}

The time reversal of pointon pair annihilation, ${[a 0] \! \oslash \!  [ 0 b]} \to {[00]}$, $a,b \in \{ \uparrow, \downarrow \}$, is pointon pair creation,
\begin{equation*}
[00] \! \oslash \!  [00] \to [ab].
\end{equation*}
The set of all possible orbital fusions, determined from the Dirac Lagrangian in Theorem \ref{allorbitals}, is therefore
\begin{equation*} \label{all orbital fusions}
\text{\footnotesize{$
{\arraycolsep=.3pt 
\begin{array}{cccccccccccc}
\xymatrix @C=-.3pc @R-2pc{
[ 0 & 0 ]\\
[ 0 & 0 ] \\
\ar@{}[rr]^(-.45){}="a"^(1.0){}="b" \ar@{-} "a";"b" &&&&\\
[ 0 & 0 ] }
&
\xymatrix @C=-.3pc @R-2pc{
[ 0 & 0 \ar@{}[ld]^(.11){}="e"^(.95){}="f" \ar@{-} "e";"f" ]\\
[ 0 & 0 ] \\
\ar@{}[rr]^(-.45){}="a"^(1.0){}="b" \ar@{-} "a";"b" &&&&\\
[ a & b ] }
&
\xymatrix @C=-.3pc @R-2pc{
[ a & 0 ]\\
[ 0 & b ] \\
\ar@{}[rr]^(-.45){}="a"^(1.0){}="b" \ar@{-} "a";"b" &&&&\\
[ a & b ] }
&
\xymatrix @C=-.3pc @R-2pc{
[ a \ar@{}[rd]^(.05){}="e"^(.9){}="f" \ar@{-} "e";"f" & 0 ]\\
[ 0 & b ] \\
\ar@{}[rr]^(-.45){}="a"^(1.0){}="b" \ar@{-} "a";"b" &&&&\\
[ 0 & 0 ] }
&
\xymatrix @C=-.3pc @R-2pc{
\textcolor{red}{\pmb{(}} a & 0 ]\\
[ 0 & b \textcolor{red}{\pmb{)}}\\
\ar@{}[rr]^(-.45){}="a"^(1.0){}="b" \ar@{-} "a";"b" &&&&\\
[ a & b ] }
&
\xymatrix @C=-.3pc @R-2pc{
\textcolor{red}{\pmb{(}} a \ar@{}[rd]^(.05){}="e"^(.9){}="f" \ar@{-} "e";"f" & 0 ]\\
[ 0 & b \textcolor{red}{\pmb{)}}\\
\ar@{}[rr]^(-.45){}="a"^(1.0){}="b" \ar@{-} "a";"b" &&&&\\
[ 0 & 0 ] }
&
\xymatrix @C=-.3pc @R-2pc{
[ \downarrow & {*} \ar@{}[ld]^(.11){}="e"^(.95){}="f" \ar@{-} "e";"f" ]\\
[ {*} & \downarrow ] \\
\ar@{}[rr]^(-.45){}="a"^(1.0){}="b" \ar@{-} "a";"b" &&&&\\
[ \downarrow & \downarrow ] }
&
\xymatrix @C=-.3pc @R-2pc{
[ \downarrow \ar@{}[rd]^(.05){}="e"^(.9){}="f" \ar@{-} "e";"f" & {*} \ar@{}[ld]^(.11){}="e"^(.95){}="f" \ar@{-} "e";"f" ]\\
[ {*} & \downarrow ] \\
\ar@{}[rr]^(-.45){}="a"^(1.0){}="b" \ar@{-} "a";"b" &&&&\\
[ 0 & 0 ] }
&
\xymatrix @C=-.3pc @R-2pc{
[ a & {*} \textcolor{red}{\pmb{)}}\ar@{}[ld]^(.11){}="e"^(.95){}="f" \ar@{-} "e";"f" \\
\textcolor{red}{\pmb{(}} {*} & b ] \\
\ar@{}[rr]^(-.45){}="a"^(1.0){}="b" \ar@{-} "a";"b" &&&&\\
[ a & b ] }
&
\xymatrix @C=-.3pc @R-2pc{
[ a \ar@{}[rd]^(.05){}="e"^(.9){}="f" \ar@{-} "e";"f"
& {*} \textcolor{red}{\pmb{)}} \ar@{}[ld]^(.11){}="e"^(.95){}="f" \ar@{-} "e";"f" \\
\textcolor{red}{\pmb{(}} {*} & b ] \\
\ar@{}[rr]^(-.45){}="a"^(1.0){}="b" \ar@{-} "a";"b" &&&&\\
[ 0 & 0 ] }
&
\xymatrix @C=-.3pc @R-2pc{
[ a & b \ar@{}[ld]^(.11){}="e"^(.95){}="f" \ar@{-} "e";"f" ]\\
[ c & d ] \\
\ar@{}[rr]^(-.45){}="a"^(1.0){}="b" \ar@{-} "a";"b" &&&&\\
[ a & d ] }
&
\xymatrix @C=-.3pc @R-2pc{
[ a \ar@{}[rd]^(.05){}="e"^(.9){}="f" \ar@{-} "e";"f"
& b \ar@{}[ld]^(.11){}="e"^(.95){}="f" \ar@{-} "e";"f" ]\\
[ c & d ] \\
\ar@{}[rr]^(-.45){}="a"^(1.0){}="b" \ar@{-} "a";"b" &&&&\\
[ 0 & 0 ] }
\end{array} }
$}}
\end{equation*}
with $a, b, c, d \in \{ \uparrow, \downarrow \}$, and $c, d$ chosen so that there is not precisely one $\uparrow$ or one $\downarrow$ pointon in the two orbitals combined.
Further, since opposite color charges cancel, we have ${\textcolor{red}{\pmb{(}} a 0 ] + [ 0 b \textcolor{red}{\pmb{)}}} = {\textcolor{red}{\pmb{(}} a b \textcolor{red}{\pmb{)}}} = {[a b]}$.

\section{Geom fusions: standard model vertices}

We call the pairs $(a_-,b_+)$ and $(b_-,a_+)$ of a geom ${[a_-a_+,b_-b_+]}$ its \textit{cross-orbitals}. 

\begin{Definition} \label{fusion def} (Geom fusions.) \rm{
Two geoms $\delta_1 = {[ \psi_{11}, \psi_{12} ]}$ and $\delta_2 = {[ \psi_{21}, \psi_{22} ]}$ may \textit{fuse} to form a geom $\delta_3$, written ${\delta_1 \! \oslash \!  \delta_2} \to \delta_3$, if $\delta_3$ results from either
\begin{itemize}
 \item[(i)] two orbital fusions ${\psi_{1i} \! \oslash \!  \psi_{2i}}$, $i \in \{ 1,2 \}$, at least one of which is of the form
\begin{equation} \label{charged fusion}
\text{\footnotesize{$
{\arraycolsep=2.9pt 
\begin{array}{c}
\xymatrix @C=-.3pc @R-2pc{
[ a & c \ar@{}[ld]^(.11){}="e"^(.95){}="f" \ar@{-} "e";"f" ]\\
[ c & b ] \\
\ar@{}[rr]^(-.45){}="a"^(1.0){}="b" \ar@{-} "a";"b" &&&&\\
[ a & b ] }
\end{array} }
$}}
\end{equation}
with $a,b \in \{ \uparrow, \downarrow \}$ and $c \in \{ 0, * \}$; or
 \item[(ii)] precisely one diagonal fusion ${\psi_{1i} \! \oslash \!  \psi_{2,i+1}}$, and $(*,0)$ is not a cross-orbital of $\delta_3$.
\end{itemize}
}\end{Definition}

In our composite model, a fusion ${\delta_1 \! \oslash \!  \delta_2} \to \delta_3$ corresponds to a trivalent interaction vertex $\delta_1 \delta_2 \delta_3$ in a Feynman diagram.
As such, fusions may be rotated in spacetime, and $4$-momentum is assumed to be conserved. 

The condition in (i) that at least one of the orbital fusions is of the form (\ref{charged fusion}), and the condition in (ii) that $\delta_3$ does not contain the cross-orbital $(*,0)$, serves to exclude certain concatenations that do not correspond to standard model vertices.
For example, the concatenations 
\begin{equation*}
\text{\footnotesize{$
{\arraycolsep=2.2pt 
\begin{array}{cccccc}
\xymatrix @C=-.3pc @R-2pc{
\gamma_a & [ a  & b, \ar@{}[ld]^(.11){}="e"^(.95){}="f" \ar@{-} "e";"f" & 0 & 0 ]\\
\gamma_a & [ a & b, & 0 & 0 ] \\
\ar@{}[rrrr]^(-.2){}="a"^(1.15){}="b" \ar@{-} "a";"b" &&&&\\
\gamma_a & [ a & b, & 0 & 0 ] }
&
\xymatrix @C=-.3pc @R-2pc{
\gamma_a & [ a \ar@{}[rd]^(.05){}="e"^(.9){}="f" \ar@{-} "e";"f" & b, \ar@{}[ld]^(.11){}="e"^(.95){}="f" \ar@{-} "e";"f"  & 0 & 0 \ar@{}[ld]^(.11){}="e"^(.95){}="f" \ar@{-} "e";"f" ]\\
\gamma_a & [ a & b, & 0 & 0 ] \\
\ar@{}[rrrr]^(-.2){}="a"^(1.15){}="b" \ar@{-} "a";"b" &&&&\\
\gamma_a & [ 0 & 0, & a & b ] }
&
\xymatrix @C=-.3pc @R-2pc{
\gamma_a & [ a  & b, \ar@{}[ld]^(.11){}="e"^(.95){}="f" \ar@{-} "e";"f" & 0 & 0 ]\\
\gamma_b & [ b & a, & 0 & 0 ] \\
\ar@{}[rrrr]^(-.2){}="a"^(1.15){}="b" \ar@{-} "a";"b" &&&&\\
Z_a & [ a & a, & 0 & 0 ] }
&
\xymatrix @C=-.3pc @R-2pc{
\gamma_a & [ a \ar@{}[rd]^(.05){}="e"^(.9){}="f" \ar@{-} "e";"f" & b, \ar@{}[ld]^(.11){}="e"^(.95){}="f" \ar@{-} "e";"f"  & 0 & 0 \ar@{}[ld]^(.11){}="e"^(.95){}="f" \ar@{-} "e";"f" ]\\
\gamma_a & [ a & b, & 0 & 0 ] \\
\ar@{}[rrrr]^(-.2){}="a"^(1.15){}="b" \ar@{-} "a";"b" &&&&\\
Z_a & [ 0 & 0, & a & a ] }
&
\xymatrix @C=-.3pc @R-2pc{
Z_a & [ a  & a, \ar@{}[ld]^(.11){}="e"^(.95){}="f" \ar@{-} "e";"f" & 0 & 0 ]\\
Z_b & [ b & b, & 0 & 0 ] \\
\ar@{}[rrrr]^(-.2){}="a"^(1.15){}="b" \ar@{-} "a";"b" &&&&\\
\gamma_a & [ a & b, & 0 & 0 ] }
&
\xymatrix @C=-.3pc @R-2pc{
Z_a & [ a \ar@{}[rd]^(.05){}="e"^(.9){}="f" \ar@{-} "e";"f" & a, \ar@{}[ld]^(.11){}="e"^(.95){}="f" \ar@{-} "e";"f"  & 0 & 0 \ar@{}[ld]^(.11){}="e"^(.95){}="f" \ar@{-} "e";"f" ]\\
Z_b & [ b & b, & 0 & 0 ] \\
\ar@{}[rrrr]^(-.2){}="a"^(1.15){}="b" \ar@{-} "a";"b" &&&&\\
\gamma_a & [ 0 & 0, & a & b ] }
\\
\addlinespace
\xymatrix @C=-.3pc @R-2pc{
Z_a & [ a \ar@{}[rd]^(.05){}="e"^(.9){}="f" \ar@{-} "e";"f" & a, \ar@{}[ld]^(.11){}="e"^(.95){}="f" \ar@{-} "e";"f"  & 0 & 0 \ar@{}[ld]^(.11){}="e"^(.95){}="f" \ar@{-} "e";"f" ]\\
Z_b & [ b & b, & 0 & 0 ] \\
\ar@{}[rrrr]^(-.2){}="a"^(1.15){}="b" \ar@{-} "a";"b" &&&&\\
Z_a & [ 0 & 0, & a & a ] }
&
\xymatrix @C=-.3pc @R-2pc{
e_a & [ 0 & 0, & a & 0 ] \\ 
\nu_e & [ {*} & \downarrow, & 0 & 0 ]\\
\ar@{}[rrrr]^(-.2){}="a"^(1.15){}="b" \ar@{-} "a";"b" &&&&\\
W^-_a & [ {*} & \downarrow, & a & 0]}
& 
\xymatrix @C=-.3pc @R-2pc{
\mu_{a} & [ b \ar@{}[rrrd]^(.05){}="c"^(.95){}="d" \ar@{-} "c";"d" & a, \ar@{}[rd]^(.05){}="e"^(.9){}="f" \ar@{-} "e";"f" & a & 0 ]\\
\nu_{\mu} & [ {*} & \downarrow, & \downarrow & \uparrow ] \\
\ar@{}[rrrr]^(-.2){}="a"^(1.15){}="b" \ar@{-} "a";"b" &&&&\\
W^-_a & [ {*} & \downarrow, & a & 0 ] } 
& 
\xymatrix @C=-.3pc @R-2pc{
\tau_{a} & [ b \ar@{}[rrrd]^(.05){}="c"^(.95){}="d" \ar@{-} "c";"d" & b, \ar@{}[rd]^(.05){}="e"^(.9){}="f" \ar@{-} "e";"f" & a & 0 ]\\
\nu_{\tau} & [ {*} & \downarrow, & \uparrow & \uparrow ] \\
\ar@{}[rrrr]^(-.2){}="a"^(1.15){}="b" \ar@{-} "a";"b" &&&&\\
W^-_a & [ {*} & \downarrow, & a & 0 ] }
&& 
\end{array}
}$}}
\end{equation*}
with $a,b \in \{a,b\}$, $a \not = b$, are not fusions.

The fermion fusions that yield the $\gamma$, $Z$, and $H$ bosons are given in Table \ref{ffZ}, and most are of fusion type (i).
The $W \! \oslash \! W \to \gamma$ and $W \! \oslash \! W \to Z$ fusions are given in Table \ref{WWgamma}, and are of both fusion types.
In the next section we will consider the lepton fusions that yield the $W$ bosons, given in Table \ref{parity}, and these are all of fusion type (ii).

\afterpage{
\setlength{\tabcolsep}{-1.4pt}
\begin{longtable}[h]{lcccccc} 
\caption{The fusions $f \! \oslash \! \bar{f} \to \gamma/Z/H$, where $f$ is a fermion. Here, $a, b \in \{ \uparrow, \downarrow \}$ with $a \not = b$.}
\endfirsthead  
\multicolumn{7}{@{}l}{\emph{(table continued from previous page)}}
\endhead
\multicolumn{7}{@{}l}{\emph{(table continued on next page)}}
\endfoot
\endlastfoot  
{\footnotesize $\ell \ell \gamma$:} &
{\footnotesize $\xymatrix @C=-.3pc @R-2pc{
e_{a} & [ a & 0, & 0 & 0 ]\\
\bar{e}_{a} & [ 0 & b, & 0 & 0 ] \\
\ar@{}[rrrr]^(-.2){}="a"^(1.15){}="b" \ar@{-} "a";"b" &&&&\\
\gamma_{a} & [ a & b, & 0 & 0 ] }$ }
&
&
{\footnotesize $\xymatrix @C=-.3pc @R-2pc{
\mu_{a} & [ a & 0, & b \ar@{}[rd]^(.05){}="e"^(.9){}="f" \ar@{-} "e";"f" & a \ar@{}[ld]^(.11){}="e"^(.95){}="f" \ar@{-} "e";"f" ]\\
\bar{\mu}_{a} & [ 0 & b, & b & a ] \\
\ar@{}[rrrr]^(-.2){}="a"^(1.15){}="b" \ar@{-} "a";"b" &&&&\\
\gamma_{a} & [ a & b, & 0 & 0 ] }$ }
&
&
{\footnotesize $\xymatrix @C=-.3pc @R-2pc{
\tau_{a} & [ a & 0, & b \ar@{}[rd]^(.05){}="e"^(.9){}="f" \ar@{-} "e";"f" & b \ar@{}[ld]^(.11){}="e"^(.95){}="f" \ar@{-} "e";"f" ]\\
\bar{\tau}_{a} & [ 0 & b, & a & a ] \\
\ar@{}[rrrr]^(-.2){}="a"^(1.15){}="b" \ar@{-} "a";"b" &&&&\\
\gamma_{a} & [ a & b, & 0 & 0 ] }$ }
&
\\
\addlinespace
{\footnotesize $qq \gamma$:} &
{\footnotesize $\xymatrix @C=-.3pc @R-2pc{
d_{a} & \textcolor{red}{\pmb{(}} a & 0, & 0 & 0 ]\\
\bar{d}_{a} & [ 0 & b, & 0 & 0 \textcolor{red}{\pmb{)}} \\
\ar@{}[rrrr]^(-.2){}="a"^(1.15){}="b" \ar@{-} "a";"b" &&&&\\
\gamma_{a} & [ a & b, & 0 & 0 ] }$ }
&
{\footnotesize $\xymatrix @C=-.3pc @R-2pc{
\bar{u}_{a} & [ a & {*}, \ar@{}[ld]^(.11){}="e"^(.95){}="f" \ar@{-} "e";"f" & 0 & 0 \textcolor{red}{\pmb{)}}\\
u_{a} & \textcolor{red}{\pmb{(}} {*} & b, & 0 & 0 ] \\
\ar@{}[rrrr]^(-.2){}="a"^(1.15){}="b" \ar@{-} "a";"b" &&&&\\
\gamma_{a} & [ a & b, & 0 & 0 ] }$ }
&
{\footnotesize $\xymatrix @C=-.3pc @R-2pc{
s_{a} & \textcolor{red}{\pmb{(}}a & 0, & b \ar@{}[rd]^(.05){}="e"^(.9){}="f" \ar@{-} "e";"f" & a \ar@{}[ld]^(.11){}="e"^(.95){}="f" \ar@{-} "e";"f" ]\\
\bar{s}_{a} & [ 0 & b, & b & a \textcolor{red}{\pmb{)}}\\
\ar@{}[rrrr]^(-.2){}="a"^(1.15){}="b" \ar@{-} "a";"b" &&&&\\
\gamma_{a} & [ a & b, & 0 & 0 ] }$ }
&
{\footnotesize $\xymatrix @C=-.3pc @R-2pc{
\bar{c}_{a} & [ a & {*}, \ar@{}[ld]^(.11){}="e"^(.95){}="f" \ar@{-} "e";"f" & b \ar@{}[rd]^(.05){}="e"^(.9){}="f" \ar@{-} "e";"f" & a \ar@{}[ld]^(.11){}="e"^(.95){}="f" \ar@{-} "e";"f" \textcolor{red}{\pmb{)}}\\
c_{a} & \textcolor{red}{\pmb{(}} {*} & b, & b & a ] \\
\ar@{}[rrrr]^(-.2){}="a"^(1.15){}="b" \ar@{-} "a";"b" &&&&\\
\gamma_{a} & [ a & b, & 0 & 0 ] }$ }
&
{\footnotesize $\xymatrix @C=-.3pc @R-2pc{
b_{a} & \textcolor{red}{\pmb{(}} a & 0, & b \ar@{}[rd]^(.05){}="e"^(.9){}="f" \ar@{-} "e";"f" & b \ar@{}[ld]^(.11){}="e"^(.95){}="f" \ar@{-} "e";"f" ]\\
\bar{b}_{a} & [ 0 & b, & a & a \textcolor{red}{\pmb{)}}\\
\ar@{}[rrrr]^(-.2){}="a"^(1.15){}="b" \ar@{-} "a";"b" &&&&\\
\gamma_{a} & [ a & b, & 0 & 0 ] }$ }
&
{\footnotesize $\xymatrix @C=-.3pc @R-2pc{
\bar{t}_{a} & [ a & {*}, \ar@{}[ld]^(.11){}="e"^(.95){}="f" \ar@{-} "e";"f", & b \ar@{}[rd]^(.05){}="e"^(.9){}="f" \ar@{-} "e";"f" & b \ar@{}[ld]^(.11){}="e"^(.95){}="f" \ar@{-} "e";"f" \textcolor{red}{\pmb{)}}\\
t_{a} & \textcolor{red}{\pmb{(}} {*} & b, & a & a ] \\
\ar@{}[rrrr]^(-.2){}="a"^(1.15){}="b" \ar@{-} "a";"b" &&&&\\
\gamma_{a} & [ a & b, & 0 & 0 ] }$ }
\\
\addlinespace
{\footnotesize $\ell \ell Z_a$:} &
{\footnotesize $\xymatrix @C=-.3pc @R-2pc{
e_{a} & [ a & 0, & 0 & 0 ]\\
\bar{e}_b & [ 0 & a, & 0 & 0 ] \\
\ar@{}[rrrr]^(-.2){}="a"^(1.15){}="b" \ar@{-} "a";"b" &&&&\\
Z_{a} & [ a & a, & 0 & 0 ] }$ }
&
{\footnotesize $\xymatrix @C=-.3pc @R-2pc{
\bar{\nu}_e & [ \downarrow & {*}, \ar@{}[ld]^(.11){}="e"^(.95){}="f" \ar@{-} "e";"f" & 0 & 0 ]\\
\nu_e & [ {*} & \downarrow, & 0 & 0 ] \\
\ar@{}[rrrr]^(-.2){}="a"^(1.15){}="b" \ar@{-} "a";"b" &&&&\\
Z_{\downarrow} & [ \downarrow & \downarrow, & 0 & 0 ] }$ }
&
{\footnotesize $\xymatrix @C=-.3pc @R-2pc{
\mu_{a} & [ a & 0, & a \ar@{}[rd]^(.05){}="e"^(.9){}="f" \ar@{-} "e";"f" & b \ar@{}[ld]^(.11){}="e"^(.95){}="f" \ar@{-} "e";"f" ]\\
\bar{\mu}_{b} & [ 0 & a, & a & b ] \\
\ar@{}[rrrr]^(-.2){}="a"^(1.15){}="b" \ar@{-} "a";"b" &&&&\\
Z_{a} & [ a & a, & 0 & 0 ] }$ }
&
{\footnotesize $\xymatrix @C=-.3pc @R-2pc{
\bar{\nu}_{\mu} & [ \downarrow & {*}, \ar@{}[ld]^(.11){}="e"^(.95){}="f" \ar@{-} "e";"f" & \uparrow \ar@{}[rd]^(.05){}="e"^(.9){}="f" \ar@{-} "e";"f" & \downarrow \ar@{}[ld]^(.11){}="e"^(.95){}="f" \ar@{-} "e";"f" ]\\
\nu_{\mu} & [ {*} & \downarrow, & \downarrow & \uparrow ] \\
\ar@{}[rrrr]^(-.2){}="a"^(1.15){}="b" \ar@{-} "a";"b" &&&&\\
Z_{\downarrow} & [ \downarrow & \downarrow, & 0 & 0 ] }$ }
&
{\footnotesize $\xymatrix @C=-.3pc @R-2pc{
\tau_{a} & [ a & 0, & b \ar@{}[rd]^(.05){}="e"^(.9){}="f" \ar@{-} "e";"f" & b \ar@{}[ld]^(.11){}="e"^(.95){}="f" \ar@{-} "e";"f" ]\\
\bar{\tau}_{b} & [ 0 & a, & b & b ] \\
\ar@{}[rrrr]^(-.2){}="a"^(1.15){}="b" \ar@{-} "a";"b" &&&&\\
Z_{a} & [ a & a, & 0 & 0 ] }$ }
&
{\footnotesize $\xymatrix @C=-.3pc @R-2pc{
\bar{\nu}_{\tau} & [ \downarrow & {*}, \ar@{}[ld]^(.11){}="e"^(.95){}="f" \ar@{-} "e";"f" & \uparrow \ar@{}[rd]^(.05){}="e"^(.9){}="f" \ar@{-} "e";"f" & \uparrow \ar@{}[ld]^(.11){}="e"^(.95){}="f" \ar@{-} "e";"f" ]\\
\nu_{\tau} & [ {*} & \downarrow, & \uparrow & \uparrow ] \\
\ar@{}[rrrr]^(-.2){}="a"^(1.15){}="b" \ar@{-} "a";"b" &&&&\\
Z_{\downarrow} & [ \downarrow & \downarrow, & 0 & 0 ] }$ }
\\
\addlinespace 
{\footnotesize $qq Z_a$:} &
{\footnotesize $\xymatrix @C=-.3pc @R-2pc{
d_{a} & \textcolor{red}{\pmb{(}} a & 0, & 0 & 0 ]\\
\bar{d}_b & [ 0 & a, & 0 & 0 \textcolor{red}{\pmb{)}}\\
\ar@{}[rrrr]^(-.2){}="a"^(1.15){}="b" \ar@{-} "a";"b" &&&&\\
Z_{a} & [ a & a, & 0 & 0 ] }$ }
&
{\footnotesize $\xymatrix @C=-.3pc @R-2pc{
\bar{u}_{a} & [a & {*}, \ar@{}[ld]^(.11){}="e"^(.95){}="f" \ar@{-} "e";"f" & 0 & 0 \textcolor{red}{\pmb{)}}\\
u_{b} & \textcolor{red}{\pmb{(}} {*} & a, & 0 & 0 ] \\
\ar@{}[rrrr]^(-.2){}="a"^(1.15){}="b" \ar@{-} "a";"b" &&&&\\
Z_{a} & [ a & a, & 0 & 0 ] }$ }
&
{\footnotesize $\xymatrix @C=-.3pc @R-2pc{
s_{a} & \textcolor{red}{\pmb{(}}a & 0, & b \ar@{}[rd]^(.05){}="e"^(.9){}="f" \ar@{-} "e";"f" & a \ar@{}[ld]^(.11){}="e"^(.95){}="f" \ar@{-} "e";"f" ]\\
\bar{s}_{b} & [ 0 & a, & a & b \textcolor{red}{\pmb{)}} \\
\ar@{}[rrrr]^(-.2){}="a"^(1.15){}="b" \ar@{-} "a";"b" &&&&\\
Z_{a} & [ a & a, & 0 & 0 ] }$ }
&
{\footnotesize $\xymatrix @C=-.3pc @R-2pc{
\bar{c}_{a} & [ a & {*}, \ar@{}[ld]^(.11){}="e"^(.95){}="f" \ar@{-} "e";"f" & b \ar@{}[rd]^(.05){}="e"^(.9){}="f" \ar@{-} "e";"f" & a \ar@{}[ld]^(.11){}="e"^(.95){}="f" \ar@{-} "e";"f" \textcolor{red}{\pmb{)}}\\
c_{b} & \textcolor{red}{\pmb{(}} {*} & a, & a & b ] \\
\ar@{}[rrrr]^(-.2){}="a"^(1.15){}="b" \ar@{-} "a";"b" &&&&\\
Z_{b} & [ a & a, & 0 & 0 ] }$ }
&
{\footnotesize $\xymatrix @C=-.3pc @R-2pc{
b_{a} & \textcolor{red}{\pmb{(}}a & 0, & b \ar@{}[rd]^(.05){}="e"^(.9){}="f" \ar@{-} "e";"f" & b \ar@{}[ld]^(.11){}="e"^(.95){}="f" \ar@{-} "e";"f" ]\\
\bar{b}_{b} & [ 0 & a, & b & b \textcolor{red}{\pmb{)}} \\
\ar@{}[rrrr]^(-.2){}="a"^(1.15){}="b" \ar@{-} "a";"b" &&&&\\
Z_{a} & [ a & a, & 0 & 0 ] }$ }
&
{\footnotesize $\xymatrix @C=-.3pc @R-2pc{
\bar{t}_{a} & [ a & {*}, \ar@{}[ld]^(.11){}="e"^(.95){}="f" \ar@{-} "e";"f" & b \ar@{}[rd]^(.05){}="e"^(.9){}="f" \ar@{-} "e";"f" & b \ar@{}[ld]^(.11){}="e"^(.95){}="f" \ar@{-} "e";"f" \textcolor{red}{\pmb{)}}\\
t_{b} & \textcolor{red}{\pmb{(}} {*} & a, & b & b ] \\
\ar@{}[rrrr]^(-.2){}="a"^(1.15){}="b" \ar@{-} "a";"b" &&&&\\
Z_{a} & [ a & a, & 0 & 0 ] }$ }
\\
\addlinespace
{\footnotesize $\ell \ell Z_0, i$:} &
{\footnotesize $\xymatrix @C=-.3pc @R-2pc{
e_{a} & [ a & 0, & 0 & 0 \ar@{}[ld]^(.11){}="e"^(.95){}="f" \ar@{-} "e";"f" ]\\
\bar{e}_{a} & [ 0 & b, & 0 & 0 ] \\
\ar@{}[rrrr]^(-.2){}="a"^(1.15){}="b" \ar@{-} "a";"b" &&&&\\
Z_0 & [ a & b, & b & a ] }$ }
&
&
{\footnotesize $\xymatrix @C=-.3pc @R-2pc{
\mu_{a} & [ a & 0, & b & a \ar@{}[ld]^(.11){}="e"^(.95){}="f" \ar@{-} "e";"f" ]\\
\bar{\mu}_{a} & [ 0 & b, & b & a ] \\
\ar@{}[rrrr]^(-.2){}="a"^(1.15){}="b" \ar@{-} "a";"b" &&&&\\
Z_0 & [ a & b, & b & a ] }$ }
&
&
{\footnotesize $\xymatrix @C=-.3pc @R-2pc{
\tau_{a} & [ a & 0, & b & b \ar@{}[ld]^(.11){}="e"^(.95){}="f" \ar@{-} "e";"f" ]\\
\bar{\tau}_{a} & [ 0 & b, & a & a ] \\
\ar@{}[rrrr]^(-.2){}="a"^(1.15){}="b" \ar@{-} "a";"b" &&&&\\
Z_0 & [ a & b, & b & a ] }$ }
&
\\
\addlinespace
{\footnotesize $qq Z_0, i$:} \! \! \!  &
{\footnotesize $\xymatrix @C=-.3pc @R-2pc{
d_{a} & \textcolor{red}{\pmb{(}}a & 0, & 0 & 0 \ar@{}[ld]^(.11){}="e"^(.95){}="f" \ar@{-} "e";"f" ]\\
\bar{d}_{a} & [ 0 & b, & 0 & 0 \textcolor{red}{\pmb{)}} \\
\ar@{}[rrrr]^(-.2){}="a"^(1.15){}="b" \ar@{-} "a";"b" &&&&\\
Z_0 & [ a & b, & b & a ] }$ }
&
{\footnotesize $\xymatrix @C=-.3pc @R-2pc{
\bar{u}_{a} & [ a & {*}, \ar@{}[ld]^(.11){}="e"^(.95){}="f" \ar@{-} "e";"f" & 0 & 0 \ar@{}[ld]^(.11){}="e"^(.95){}="f" \ar@{-} "e";"f" \textcolor{red}{\pmb{)}}\\
u_{a} & \textcolor{red}{\pmb{(}} {*} & b, & 0 & 0 ]\\
\ar@{}[rrrr]^(-.2){}="a"^(1.15){}="b" \ar@{-} "a";"b" &&&&\\
Z_0 & [ a & b, & b & a ] }$ }
&
{\footnotesize $\xymatrix @C=-.3pc @R-2pc{
s_{a} & \textcolor{red}{\pmb{(}}a & 0, & b & a \ar@{}[ld]^(.11){}="e"^(.95){}="f" \ar@{-} "e";"f" ]\\
\bar{s}_{a} & [ 0 & b, & b & a \textcolor{red}{\pmb{)}}\\
\ar@{}[rrrr]^(-.2){}="a"^(1.15){}="b" \ar@{-} "a";"b" &&&&\\
Z_0 & [ a & b, & b & a ] }$ }
&
{\footnotesize $\xymatrix @C=-.3pc @R-2pc{
\bar{c}_{a} & [ a & {*}, \ar@{}[ld]^(.11){}="e"^(.95){}="f" \ar@{-} "e";"f" & b & a \ar@{}[ld]^(.11){}="e"^(.95){}="f" \ar@{-} "e";"f" \textcolor{red}{\pmb{)}}\\
c_{a} & \textcolor{red}{\pmb{(}} {*} & a, & a & b ] \\
\ar@{}[rrrr]^(-.2){}="a"^(1.15){}="b" \ar@{-} "a";"b" &&&&\\
Z_0 & [ a & b, & b & a ] }$ }
&
{\footnotesize $\xymatrix @C=-.3pc @R-2pc{
b_{a} & \textcolor{red}{\pmb{(}}a & 0, & b & b \ar@{}[ld]^(.11){}="e"^(.95){}="f" \ar@{-} "e";"f" ]\\
\bar{b}_{a} & [ 0 & b, & a & a \textcolor{red}{\pmb{)}} \\
\ar@{}[rrrr]^(-.2){}="a"^(1.15){}="b" \ar@{-} "a";"b" &&&&\\
Z_0 & [ a & b, & b & a ] }$ }
&
{\footnotesize $\xymatrix @C=-.3pc @R-2pc{
\bar{t}_{a} & [ a & {*}, \ar@{}[ld]^(.11){}="e"^(.95){}="f" \ar@{-} "e";"f" & b & b \ar@{}[ld]^(.11){}="e"^(.95){}="f" \ar@{-} "e";"f" \textcolor{red}{\pmb{)}}\\
t_{a} & \textcolor{red}{\pmb{(}} {*} & b, & a & a ] \\
\ar@{}[rrrr]^(-.2){}="a"^(1.15){}="b" \ar@{-} "a";"b" &&&&\\
Z_0 & [ a & b, & b & a ] }$ }
\\
\addlinespace
{\footnotesize $\ell \ell Z_0, ii$:}  &
{\footnotesize $\xymatrix @C=-.3pc @R-2pc{
e_{a} & [ a & 0, & 0 \ar@{}[ld]^(.11){}="e"^(.95){}="f" \ar@{-} "e";"f" & 0 ]\\
\bar{e}_{b} & [ 0 & 0, & 0 & a ] \\
\ar@{}[rrrr]^(-.2){}="a"^(1.15){}="b" \ar@{-} "a";"b" &&&&\\
Z_0 & [ a & b, & b & a ] }$ }
&
&
{\footnotesize $\xymatrix @C=-.3pc @R-2pc{
\mu_{a} & [ a & 0, & b & a \ar@{}[llld]^(.05){}="e"^(.95){}="f" \ar@{-} "e";"f" ]\\
\bar{\mu}_{b} & [ a & b, & 0 & a ] \\
\ar@{}[rrrr]^(-.2){}="a"^(1.15){}="b" \ar@{-} "a";"b" &&&&\\
Z_0 & [ a & b, & b & a ] }$ }
&
&
{\footnotesize $\xymatrix @C=-.3pc @R-2pc{
\tau_{a} & [ a & 0, & b & b  \ar@{}[llld]^(.05){}="e"^(.95){}="f" \ar@{-} "e";"f" ]\\
\bar{\tau}_{b} & [ b & b, & 0 & a ] \\
\ar@{}[rrrr]^(-.2){}="a"^(1.15){}="b" \ar@{-} "a";"b" &&&&\\
Z_0 & [ a & b, & b & a ] }$ }
&
\\
\addlinespace
{\footnotesize $qq Z_0, ii$:} &
{\footnotesize $\xymatrix @C=-.3pc @R-2pc{
d_{a} & \textcolor{red}{\pmb{(}}a & 0, & 0 \ar@{}[ld]^(.11){}="e"^(.95){}="f" \ar@{-} "e";"f" & 0 ]\\
\bar{d}_{b} & [ 0 & 0, & 0 & a\textcolor{red}{\pmb{)}} \\
\ar@{}[rrrr]^(-.2){}="a"^(1.15){}="b" \ar@{-} "a";"b" &&&&\\
Z_0 & [ a & b, & b & a ] }$ }
&
&
{\footnotesize $\xymatrix @C=-.3pc @R-2pc{
s_{a} & \textcolor{red}{\pmb{(}}a & 0, & b & a \ar@{}[llld]^(.05){}="e"^(.95){}="f" \ar@{-} "e";"f" ]\\
\bar{s}_{b} & [ a & b, & 0 & a\textcolor{red}{\pmb{)}} \\
\ar@{}[rrrr]^(-.2){}="a"^(1.15){}="b" \ar@{-} "a";"b" &&&&\\
Z_0 & [ a & b, & b & a ] }$ }
&
&
{\footnotesize $\xymatrix @C=-.3pc @R-2pc{
b_{a} & \textcolor{red}{\pmb{(}}a & 0, & b & b  \ar@{}[llld]^(.05){}="e"^(.95){}="f" \ar@{-} "e";"f" ]\\
\bar{b}_{b} & [ b & b, & 0 & a\textcolor{red}{\pmb{)}} \\
\ar@{}[rrrr]^(-.2){}="a"^(1.15){}="b" \ar@{-} "a";"b" &&&&\\
Z_0 & [ a & b, & b & a ] }$ }
&
\\
\addlinespace
{\footnotesize $\ell \ell H, i$:} &
{\footnotesize $\xymatrix @C=-.3pc @R-2pc{
e_{a} & [ a & 0, & 0 & 0 \ar@{}[ld]^(.11){}="e"^(.95){}="f" \ar@{-} "e";"f" ]\\
\bar{e}_{b} & [ 0 & a, & 0 & 0 ] \\
\ar@{}[rrrr]^(-.2){}="a"^(1.15){}="b" \ar@{-} "a";"b" &&&&\\
H & [ a & a, & b & b ] }$ }
&
{\footnotesize $\xymatrix @C=-.3pc @R-2pc{
\bar{\nu}_e & [ \downarrow & {*}, \ar@{}[ld]^(.11){}="e"^(.95){}="f" \ar@{-} "e";"f" & 0 & 0 \ar@{}[ld]^(.11){}="e"^(.95){}="f" \ar@{-} "e";"f" ]\\
\nu_e & [ {*} & \downarrow, & 0 & 0 ] \\
\ar@{}[rrrr]^(-.2){}="a"^(1.15){}="b" \ar@{-} "a";"b" &&&&\\
H & [ \downarrow & \downarrow, & \uparrow & \uparrow ] }$ }
&
{\footnotesize $\xymatrix @C=-.3pc @R-2pc{
\mu_a & [ a & 0, & b & a \ar@{}[ld]^(.11){}="e"^(.95){}="f" \ar@{-} "e";"f" ]\\
\bar{\mu}_b & [ 0 & a, & a & b ] \\
\ar@{}[rrrr]^(-.2){}="a"^(1.15){}="b" \ar@{-} "a";"b" &&&&\\
H & [ a & a, & b & b ] }$ }
&
{\footnotesize $\xymatrix @C=-.3pc @R-2pc{
\bar{\nu}_{\mu} & [ \downarrow & {*}, \ar@{}[ld]^(.11){}="e"^(.95){}="f" \ar@{-} "e";"f" & \uparrow & \downarrow \ar@{}[ld]^(.11){}="e"^(.95){}="f" \ar@{-} "e";"f" ]\\
\nu_{\mu} & [ {*} & \downarrow, & \downarrow & \uparrow ] \\
\ar@{}[rrrr]^(-.2){}="a"^(1.15){}="b" \ar@{-} "a";"b" &&&&\\
H & [ \downarrow & \downarrow, & \uparrow & \uparrow ] }$ } 
&
{\footnotesize $\xymatrix @C=-.3pc @R-2pc{
\tau_{a} & [ a & 0, & b & b \ar@{}[ld]^(.11){}="e"^(.95){}="f" \ar@{-} "e";"f" ]\\
\bar{\tau}_{b} & [ 0 & a, & b & b ] \\
\ar@{}[rrrr]^(-.2){}="a"^(1.15){}="b" \ar@{-} "a";"b" &&&&\\
H & [ a & a, & b & b ] }$ } 
&
{\footnotesize $\xymatrix @C=-.3pc @R-2pc{
\bar{\nu}_{\tau} & [ \downarrow & {*}, \ar@{}[ld]^(.11){}="e"^(.95){}="f" \ar@{-} "e";"f" & \uparrow & \uparrow \ar@{}[ld]^(.11){}="e"^(.95){}="f" \ar@{-} "e";"f" ]\\
\nu_{\tau} & [{*} & \downarrow, & \uparrow & \uparrow ] \\
\ar@{}[rrrr]^(-.2){}="a"^(1.15){}="b" \ar@{-} "a";"b" &&&&\\
H & [ \downarrow & \downarrow, & \uparrow & \uparrow ] }$ }
\\
\addlinespace
{\footnotesize $qqH, i$:} &
{\footnotesize $\xymatrix @C=-.3pc @R-2pc{
d_{a} & \textcolor{red}{\pmb{(}}a & 0, & 0 & 0 \ar@{}[ld]^(.11){}="e"^(.95){}="f" \ar@{-} "e";"f" ]\\
\bar{d}_{b} & [ 0 & a, & 0 & 0 \textcolor{red}{\pmb{)}} \\
\ar@{}[rrrr]^(-.2){}="a"^(1.15){}="b" \ar@{-} "a";"b" &&&&\\
H & [ a & a, & b & b ] }$ } 
&
{\footnotesize $\xymatrix @C=-.3pc @R-2pc{
\bar{u}_{a} & [ a & {*}, \ar@{}[ld]^(.11){}="e"^(.95){}="f" \ar@{-} "e";"f" & 0 & 0 \ar@{}[ld]^(.11){}="e"^(.95){}="f" \ar@{-} "e";"f" \textcolor{red}{\pmb{)}}\\
u_{a} & \textcolor{red}{\pmb{(}} {*} & a, & 0 & 0 ] \\
\ar@{}[rrrr]^(-.2){}="a"^(1.15){}="b" \ar@{-} "a";"b" &&&&\\
H & [ a & a, & b & b ] }$ }
&
{\footnotesize $\xymatrix @C=-.3pc @R-2pc{
s_{a} & \textcolor{red}{\pmb{(}}a & 0, & b & a \ar@{}[ld]^(.11){}="e"^(.95){}="f" \ar@{-} "e";"f" ]\\
\bar{s}_{b} & [ 0 & a, & a & b \textcolor{red}{\pmb{)}} \\
\ar@{}[rrrr]^(-.2){}="a"^(1.15){}="b" \ar@{-} "a";"b" &&&&\\
H & [ a & a, & b & b ] }$ }
&
{\footnotesize $\xymatrix @C=-.3pc @R-2pc{
\bar{c}_{a} & [ a & {*}, \ar@{}[ld]^(.11){}="e"^(.95){}="f" \ar@{-} "e";"f" & b & a \ar@{}[ld]^(.11){}="e"^(.95){}="f" \ar@{-} "e";"f" \textcolor{red}{\pmb{)}} \\
c_{a} & \textcolor{red}{\pmb{(}} {*} & a, & a & b ]\\
\ar@{}[rrrr]^(-.2){}="a"^(1.15){}="b" \ar@{-} "a";"b" &&&&\\
H & [ a & a, & b & b ] }$ }
&
{\footnotesize $\xymatrix @C=-.3pc @R-2pc{
b_{a} & \textcolor{red}{\pmb{(}}a & 0, & b & b \ar@{}[ld]^(.11){}="e"^(.95){}="f" \ar@{-} "e";"f" ]\\
\bar{b}_{b} & [ 0 &  a, & b & b \textcolor{red}{\pmb{)}} \\
\ar@{}[rrrr]^(-.2){}="a"^(1.15){}="b" \ar@{-} "a";"b" &&&&\\
H & [ a & a, & b & b ] }$ }
&
{\footnotesize $\xymatrix @C=-.3pc @R-2pc{
\bar{t}_{a} & [ a & {*}, \ar@{}[ld]^(.11){}="e"^(.95){}="f" \ar@{-} "e";"f" & b & b \ar@{}[ld]^(.11){}="e"^(.95){}="f" \ar@{-} "e";"f" \textcolor{red}{\pmb{)}}\\
t_{b} & \textcolor{red}{\pmb{(}} {*} & a, & b & b ] \\
\ar@{}[rrrr]^(-.2){}="a"^(1.15){}="b" \ar@{-} "a";"b" &&&&\\
H & [ a & a, & b & b ] }$ }
\\
\addlinespace
{\footnotesize $\ell \ell H, ii$:} &
{\footnotesize $\xymatrix @C=-.3pc @R-2pc{
e_{a} & [ a & 0, & 0  \ar@{}[ld]^(.11){}="e"^(.95){}="f" \ar@{-} "e";"f" & 0 ]\\
\bar{e}_{a} & [ 0 & 0, & 0 & b ] \\
\ar@{}[rrrr]^(-.2){}="a"^(1.15){}="b" \ar@{-} "a";"b" &&&&\\
H & [ a & a, & b & b ] }$ }
&
&
{\footnotesize $\xymatrix @C=-.3pc @R-2pc{
\mu_a & [ a & 0, & b & a \ar@{}[llld]^(.05){}="e"^(.95){}="f" \ar@{-} "e";"f" ]\\
\bar{\mu}_a & [ b & a, & 0 & b ] \\
\ar@{}[rrrr]^(-.2){}="a"^(1.15){}="b" \ar@{-} "a";"b" &&&&\\
H & [ a & a, & b & b ] }$ }
&
&
{\footnotesize $\xymatrix @C=-.3pc @R-2pc{
\tau_{a} & [ a & 0, & b & b \ar@{}[llld]^(.05){}="e"^(.95){}="f" \ar@{-} "e";"f" ]\\
\bar{\tau}_{a} & [ a & a, & 0 & b ] \\
\ar@{}[rrrr]^(-.2){}="a"^(1.15){}="b" \ar@{-} "a";"b" &&&&\\
H & [ a & a, & b & b ] }$ }
&
\\
\addlinespace
{\footnotesize $qq H, ii$:} &
{\footnotesize $\xymatrix @C=-.3pc @R-2pc{
d_{a} & \textcolor{red}{\pmb{(}}a & 0, & 0  \ar@{}[ld]^(.11){}="e"^(.95){}="f" \ar@{-} "e";"f" & 0 ]\\
\bar{d}_{a} & [ 0 & 0, & 0 & b\textcolor{red}{\pmb{)}} \\
\ar@{}[rrrr]^(-.2){}="a"^(1.15){}="b" \ar@{-} "a";"b" &&&&\\
H & [ a & a, & b & b ] }$ }
&
&
{\footnotesize $\xymatrix @C=-.3pc @R-2pc{
s_a & \textcolor{red}{\pmb{(}}a & 0, & b & a \ar@{}[llld]^(.05){}="e"^(.95){}="f" \ar@{-} "e";"f" ]\\
\bar{s}_a & [ b & a, & 0 & b\textcolor{red}{\pmb{)}} \\
\ar@{}[rrrr]^(-.2){}="a"^(1.15){}="b" \ar@{-} "a";"b" &&&&\\
H & [ a & a, & b & b ] }$ }
&
&
{\footnotesize $\xymatrix @C=-.3pc @R-2pc{
b_{a} & \textcolor{red}{\pmb{(}}a & 0, & b & b \ar@{}[llld]^(.05){}="e"^(.95){}="f" \ar@{-} "e";"f" ]\\
\bar{b}_{a} & [ a & a, & 0 & b\textcolor{red}{\pmb{)}} \\
\ar@{}[rrrr]^(-.2){}="a"^(1.15){}="b" \ar@{-} "a";"b" &&&&\\
H & [ a & a, & b & b ] }$ }
&
\label{ffZ}
\end{longtable} }

\afterpage{
\setlength{\tabcolsep}{2.2pt}
\begin{longtable}[h]{cccccc} 
\caption{The three-boson vertices.}
\endfirsthead  
\multicolumn{6}{@{}l}{\emph{(table continued from previous page)}}
\endhead
\multicolumn{6}{@{}l}{\emph{(table continued on next page)}}
\endfoot
\endlastfoot  
{\footnotesize SM:} &
{\footnotesize $\xymatrix @C=-.3pc @R-2pc{
W^-_{\uparrow} & [ \uparrow & 0, & {*}  \ar@{}[rd]^(.05){}="e"^(.9){}="f" \ar@{-} "e";"f" & \downarrow \ar@{}[ld]^(.11){}="e"^(.95){}="f" \ar@{-} "e";"f" ]\\
\text{\footnotesize{$W^+_{\uparrow}$}} & [ 0 & \downarrow, & \downarrow & {*} ] \\
\ar@{}[rrrr]^(-.2){}="a"^(1.15){}="b" \ar@{-} "a";"b" &&&&\\
\gamma_{\uparrow} & [ \uparrow & \downarrow, & 0 & 0 ] }$}
&
{\footnotesize $\xymatrix @C=-.3pc @R-2pc{
W^-_{\downarrow} & [ \downarrow & 0, & {*}  \ar@{}[rd]^(.05){}="e"^(.9){}="f" \ar@{-} "e";"f" & \downarrow \ar@{}[ld]^(.11){}="e"^(.95){}="f" \ar@{-} "e";"f" ]\\
W^+_{\downarrow} & [ 0 & \uparrow, & \downarrow & {*} ] \\
\ar@{}[rrrr]^(-.2){}="a"^(1.15){}="b" \ar@{-} "a";"b" &&&&\\
\gamma_{\downarrow} & [ \downarrow & \uparrow, & 0 & 0 ] }$}
&
{\footnotesize $\xymatrix @C=-.3pc @R-2pc{
W^-_{\uparrow} & [ \uparrow & 0, & {*}  \ar@{}[rd]^(.05){}="e"^(.9){}="f" \ar@{-} "e";"f" & \downarrow \ar@{}[ld]^(.11){}="e"^(.95){}="f" \ar@{-} "e";"f" ]\\
W^+_{\downarrow} & [ 0 & \uparrow, & \downarrow & {*} ] \\
\ar@{}[rrrr]^(-.2){}="a"^(1.15){}="b" \ar@{-} "a";"b" &&&&\\
Z_{\uparrow} & [ \uparrow & \uparrow, & 0 & 0 ] }$}
&
{\footnotesize $\xymatrix @C=-.3pc @R-2pc{
W^-_{\downarrow} & [ \downarrow & 0, & {*}  \ar@{}[rd]^(.05){}="e"^(.9){}="f" \ar@{-} "e";"f" & \downarrow \ar@{}[ld]^(.11){}="e"^(.95){}="f" \ar@{-} "e";"f" ]\\
W^+_{\uparrow} & [ 0 & \downarrow, & \downarrow & {*} ] \\
\ar@{}[rrrr]^(-.2){}="a"^(1.15){}="b" \ar@{-} "a";"b" &&&&\\
Z_{\downarrow} & [ \downarrow & \downarrow, & 0 & 0 ] }$}
&
{\footnotesize $\xymatrix @C=-.3pc @R-2pc{
W^-_0 & [ \uparrow & 0, & \downarrow & {*} \ar@{}[llld]^(.05){}="e"^(.95){}="f" \ar@{-} "e";"f" ]\\
W^+_0 & [ {*} & \downarrow, & 0 & \uparrow ] \\
\ar@{}[rrrr]^(-.2){}="a"^(1.15){}="b" \ar@{-} "a";"b" &&&&\\
Z_0 & [ \uparrow & \downarrow, & \downarrow & \uparrow ] }$} 
\\
\addlinespace
&
{\footnotesize $\xymatrix @C=-.3pc @R-2pc{
W^-_{\uparrow} & [ \uparrow & 0, & {*}  \ar@{}[rd]^(.05){}="e"^(.9){}="f" \ar@{-} "e";"f" & \downarrow ]\\
\text{\footnotesize{$W^+_{\downarrow}$}} & [ 0 & \uparrow, & \downarrow & {*} ] \\
\ar@{}[rrrr]^(-.2){}="a"^(1.15){}="b" \ar@{-} "a";"b" &&&&\\
H & [ \uparrow & \uparrow, & \downarrow & \downarrow ] }$}
&
{\footnotesize $\xymatrix @C=-.3pc @R-2pc{
W^-_0 & [ \uparrow & 0, & \downarrow & {*} \ar@{}[ld]^(.11){}="e"^(.95){}="f" \ar@{-} "e";"f" ]\\
W^+_0 & [ 0 & \uparrow, & {*} & \downarrow ] \\
\ar@{}[rrrr]^(-.2){}="a"^(1.15){}="b" \ar@{-} "a";"b" &&&&\\
H & [ \uparrow & \uparrow, & \downarrow & \downarrow ] }$}
& 
{\footnotesize $\xymatrix @C=-.3pc @R-2pc{
Z_{\uparrow} & [ \uparrow & \uparrow, & 0 & 0]\\
Z_{\downarrow} & [ 0 & 0, & \downarrow & \downarrow ] \\
\ar@{}[rrrr]^(-.2){}="a"^(1.15){}="b" \ar@{-} "a";"b" &&&&\\
H & [ \uparrow & \uparrow, & \downarrow & \downarrow ] } $}
&
{\footnotesize $\xymatrix @C=-.3pc @R-2pc{
H & [ \uparrow \ar@{}[rrrd]^(.05){}="c"^(.95){}="d" \ar@{-} "c";"d" & \uparrow, \ar@{}[rd]^(.05){}="e"^(.9){}="f" \ar@{-} "e";"f" & \downarrow & \downarrow ]\\
H & [ \uparrow & \uparrow, & \downarrow & \downarrow ] \\
\ar@{}[rrrr]^(-.2){}="a"^(1.15){}="b" \ar@{-} "a";"b" &&&&\\
H & [ \uparrow & \uparrow, & \downarrow & \downarrow ] } $}
&
\\
\addlinespace
{\footnotesize non-SM:} &
{\footnotesize $\xymatrix @C=-.3pc @R-2pc{
\gamma_{\uparrow} & [ \uparrow & \downarrow, & 0 & 0 ] \\ 
\gamma_{\downarrow} & [ 0 & 0, & \downarrow & \uparrow ] \\
\ar@{}[rrrr]^(-.2){}="a"^(1.15){}="b" \ar@{-} "a";"b" &&&&\\
Z_0 & [ \uparrow & \downarrow, & \downarrow & \uparrow ]} $ }
&
{\footnotesize $\xymatrix @C=-.3pc @R-2pc{
Z_0 & [ \uparrow \ar@{}[rrrd]^(.05){}="c"^(.95){}="d" \ar@{-} "c";"d" & \downarrow,  \ar@{}[rd]^(.05){}="e"^(.9){}="f" \ar@{-} "e";"f" & \downarrow & \uparrow ] \\
Z_0 & [ \uparrow & \downarrow, & \downarrow & \uparrow ] \\
\ar@{}[rrrr]^(-.2){}="a"^(1.15){}="b" \ar@{-} "a";"b" &&&&\\
Z_0 & [ \uparrow & \downarrow, & \downarrow & \uparrow ]}$ }
&
&
&
\label{WWgamma}
\end{longtable} }

\section{A derivation of electroweak parity violation}

In the standard model, leptons and quarks transform as left-handed doublets and right-handed singlets under $\operatorname{SU}(2)$:
\begin{equation*}
\begin{array}{cccccc}
\left( \begin{matrix} \nu_e \\ e \end{matrix} \right)_{\! \! \! L} & \begin{pmatrix} \nu_{\mu} \\ \mu \end{pmatrix}_{\! \! \! L} & \begin{pmatrix} \nu_{\tau} \\ \tau \end{pmatrix}_{\! \! \! L} &
\left( \begin{matrix} u \\ d \end{matrix} \right)_{\! \! \! L} & \left( \begin{matrix} c \\ s \end{matrix} \right)_{\! \! \! L} & \left( \begin{matrix} t \\ b \end{matrix} \right)_{\! \! \! L}\\
e_R & \mu_R & \tau_R & u_R & c_R & t_R \\
&&& d_R & s_R & b_R
\end{array}
\end{equation*}
On an internal spacetime, left- and right-handed chirality is replaced by geom spin states; see \cite[Sections 2, 3]{B1}.
For example, the chiral states $e_L$, $e_R$ are replaced by the spin states $e_{\uparrow}$, $e_{\downarrow}$.

\begin{Theorem} \label{parity violation} \ 
\begin{itemize}
\item There are fusions
\begin{equation} \label{electroweak}
\ell_{\uparrow} \! \oslash \!  \bar{\nu}_{\ell} \to W^-_0, \ \ \ \ \bar{\ell}_{\downarrow} \! \oslash \!  \nu_{\ell} \to W^+_0, \ \ \ \ d_{\uparrow} \! \oslash \!  \bar{u}_{\uparrow} \to W^-_0, \ \ \ \ \bar{d}_{\downarrow} \! \oslash \!  u_{\downarrow} \to W^+_0,
\end{equation}
where $\ell$ is a charged lepton, $u$ is an up-type quark, and $d$ is a down-type quark.
\item The concatenations (\ref{electroweak}) with all other spin states $\uparrow$, $\downarrow$ are not fusions.
\end{itemize}
In particular, the composite model yields electroweak parity violation for both leptons and quarks.
\end{Theorem}

\afterpage{
\setlength{\tabcolsep}{2.2pt}
\begin{longtable}[h]{cccccc} 
\caption{The $W^{\pm}$ fusions that exhibit parity violation.} 
\endfirsthead  
\multicolumn{6}{@{}l}{(\emph{table continued from previous page})}
\endhead
\multicolumn{6}{@{}l}{(\emph{table continued on next page})}
\endfoot
\endlastfoot  
{\footnotesize $\xymatrix @C=-.3pc @R-2pc{
e_{\uparrow} & [ 0 & 0, & \uparrow & 0 ]\\
\bar{\nu}_{e} & [ \downarrow & {*}, & 0 & 0 ] \\
\ar@{}[rrrr]^(-.2){}="a"^(1.15){}="b" \ar@{-} "a";"b" &&&&\\
W^-_0 & [ \downarrow & {*}, & \uparrow & 0 ] } $}
& 
{\footnotesize $\xymatrix @C=-.3pc @R-2pc{
\bar{e}_{\downarrow} & [ 0 & 0, & 0 & \uparrow ]\\
\nu_{e} & [ {*} & \downarrow, & 0 & 0 ] \\
\ar@{}[rrrr]^(-.2){}="a"^(1.15){}="b" \ar@{-} "a";"b" &&&&\\
W^+_0 & [ {*} & \downarrow, & 0 & \uparrow ] } $}
&
{\footnotesize $\xymatrix @C=-.3pc @R-2pc{
\mu_{\uparrow} & [ \downarrow \ar@{}[rrrd]^(.05){}="c"^(.95){}="d" \ar@{-} "c";"d" & \uparrow, \ar@{}[rd]^(.05){}="e"^(.9){}="f" \ar@{-} "e";"f" & \uparrow & 0 ]\\
\bar{\nu}_{\tau} & [ \downarrow & {*}, & \uparrow & \downarrow ] \\
\ar@{}[rrrr]^(-.2){}="a"^(1.15){}="b" \ar@{-} "a";"b" &&&&\\
W^-_0 & [ \downarrow & {*}, & \uparrow & 0 ] } $}
& 
{\footnotesize $\xymatrix @C=-.3pc @R-2pc{
\bar{\mu}_{\downarrow} & [ \uparrow \ar@{}[rrrd]^(.05){}="c"^(.95){}="d" \ar@{-} "c";"d" & \downarrow, \ar@{}[rd]^(.05){}="e"^(.9){}="f" \ar@{-} "e";"f" & 0 & \uparrow ]\\
\nu_{\mu} & [ {*} & \downarrow, & \downarrow & \uparrow ] \\
\ar@{}[rrrr]^(-.2){}="a"^(1.15){}="b" \ar@{-} "a";"b" &&&&\\
W^+_0 & [ {*} & \downarrow, & 0 & \uparrow ] } $}
&
{\footnotesize $\xymatrix @C=-.3pc @R-2pc{
\tau_{\uparrow} & [ \downarrow \ar@{}[rrrd]^(.05){}="c"^(.95){}="d" \ar@{-} "c";"d" & \downarrow, \ar@{}[rd]^(.05){}="e"^(.9){}="f" \ar@{-} "e";"f" & \uparrow & 0 ]\\
\bar{\nu}_{\tau} & [ \downarrow & {*}, & \uparrow & \uparrow ] \\
\ar@{}[rrrr]^(-.2){}="a"^(1.15){}="b" \ar@{-} "a";"b" &&&&\\
W^-_0 & [ \downarrow & {*}, & \uparrow & 0 ] } $}
& 
{\footnotesize $\xymatrix @C=-.3pc @R-2pc{
\bar{\tau}_{\downarrow} & [ \downarrow \ar@{}[rrrd]^(.05){}="c"^(.95){}="d" \ar@{-} "c";"d" & \downarrow, \ar@{}[rd]^(.05){}="e"^(.9){}="f" \ar@{-} "e";"f" & 0 & \uparrow ]\\
\nu_{\tau} & [ {*} & \downarrow, & \uparrow & \uparrow ] \\
\ar@{}[rrrr]^(-.2){}="a"^(1.15){}="b" \ar@{-} "a";"b" &&&&\\
W^+_0 & [ {*} & \downarrow, & 0 & \uparrow ] } $}
\\
\addlinespace
{\footnotesize $\xymatrix @C=-.3pc @R-2pc{
d_{\uparrow} & \textcolor{red}{\pmb{(}} 0 & 0, & \uparrow & 0 ]\\
\bar{u}_{\uparrow} & [ \downarrow & {*}, & 0 & 0 \textcolor{red}{\pmb{)}} \\
\ar@{}[rrrr]^(-.2){}="a"^(1.15){}="b" \ar@{-} "a";"b" &&&&\\
W^-_0 & [ \downarrow & {*}, & \uparrow & 0 ] } $}
& 
{\footnotesize $\xymatrix @C=-.3pc @R-2pc{
\bar{d}_{\downarrow} & [ 0 & 0, & 0 & \uparrow \! \textcolor{red}{\pmb{)}}\\
u_{\downarrow} & \textcolor{red}{\pmb{(}} {*} & \downarrow, & 0 & 0 ] \\
\ar@{}[rrrr]^(-.2){}="a"^(1.15){}="b" \ar@{-} "a";"b" &&&&\\
W^+_0 & [ {*} & \downarrow, & 0 & \uparrow ] } $}
&
{\footnotesize $\xymatrix @C=-.3pc @R-2pc{
s_{\uparrow} & \textcolor{red}{\pmb{(}} \! \downarrow \ar@{}[rrrd]^(.05){}="c"^(.95){}="d" \ar@{-} "c";"d" & \uparrow, \ar@{}[rd]^(.05){}="e"^(.9){}="f" \ar@{-} "e";"f" & \uparrow & 0 ]\\
\bar{c}_{\uparrow} & [ \downarrow & {*}, & \uparrow & \downarrow \! \textcolor{red}{\pmb{)}} \\
\ar@{}[rrrr]^(-.2){}="a"^(1.15){}="b" \ar@{-} "a";"b" &&&&\\
W^-_0 & [ \downarrow & {*}, & \uparrow & 0 ] } $}
& 
{\footnotesize $\xymatrix @C=-.3pc @R-2pc{
\bar{s}_{\downarrow} & [ \uparrow \ar@{}[rrrd]^(.05){}="c"^(.95){}="d" \ar@{-} "c";"d" & \downarrow, \ar@{}[rd]^(.05){}="e"^(.9){}="f" \ar@{-} "e";"f" & 0 & \uparrow \! \textcolor{red}{\pmb{)}}\\
c_{\downarrow} & \textcolor{red}{\pmb{(}} {*} & \downarrow, & \downarrow & \uparrow ] \\
\ar@{}[rrrr]^(-.2){}="a"^(1.15){}="b" \ar@{-} "a";"b" &&&&\\
W^+_0 & [ {*} & \downarrow, & 0 & \uparrow ] } $}
&
{\footnotesize $\xymatrix @C=-.3pc @R-2pc{
b_{\uparrow} & \textcolor{red}{\pmb{(}} \! \downarrow \ar@{}[rrrd]^(.05){}="c"^(.95){}="d" \ar@{-} "c";"d" & \downarrow, \ar@{}[rd]^(.05){}="e"^(.9){}="f" \ar@{-} "e";"f" & \uparrow & 0 ]\\
\bar{t}_{\uparrow} & [ \downarrow & {*}, & \uparrow & \uparrow \! \textcolor{red}{\pmb{)}} \\
\ar@{}[rrrr]^(-.2){}="a"^(1.15){}="b" \ar@{-} "a";"b" &&&&\\
W^-_0 & [ \downarrow & {*}, & \uparrow & 0 ] } $}
& 
{\footnotesize $\xymatrix @C=-.3pc @R-2pc{
\bar{b}_{\downarrow} & [ \downarrow \ar@{}[rrrd]^(.05){}="c"^(.95){}="d" \ar@{-} "c";"d" & \downarrow, \ar@{}[rd]^(.05){}="e"^(.9){}="f" \ar@{-} "e";"f" & 0 & \uparrow \! \textcolor{red}{\pmb{)}}\\
t_{\downarrow} & \textcolor{red}{\pmb{(}} {*} & \downarrow, & \uparrow & \uparrow ] \\
\ar@{}[rrrr]^(-.2){}="a"^(1.15){}="b" \ar@{-} "a";"b" &&&&\\
W^+_0 & [ {*} & \downarrow, & 0 & \uparrow ] } $}
\label{parity}
\end{longtable} }

\begin{proof}
The concatenations given in Table \ref{parity} are fusions, and so are allowed.
However, the concatenations
\begin{equation*}
{\arraycolsep=2.2pt
\text{\footnotesize{$
\begin{array}{cccccc} 
\xymatrix @C=-.3pc @R-2pc{
e_{\downarrow} & [ 0 & 0, & \downarrow & 0 ]\\
\bar{\nu}_{e} & [ \downarrow & {*}, & 0 & 0 ] \\
\ar@{}[rrrr]^(-.2){}="a"^(1.15){}="b" \ar@{-} "a";"b" &&&&\\
 & [ \downarrow & {*}, & \downarrow & 0 ] } 
& 
\xymatrix @C=-.3pc @R-2pc{
\bar{e}_{\uparrow} & [ 0 & 0, & 0 & \downarrow ]\\
\nu_{e} & [ {*} & \downarrow, & 0 & 0 ] \\
\ar@{}[rrrr]^(-.2){}="a"^(1.15){}="b" \ar@{-} "a";"b" &&&&\\
 & [ {*} & \downarrow, & 0 & \downarrow ] } 
&
\xymatrix @C=-.3pc @R-2pc{
\mu_{\downarrow} & [ \uparrow \ar@{}[rrrd]^(.05){}="c"^(.95){}="d" \ar@{-} "c";"d" & \downarrow, \ar@{}[rd]^(.05){}="e"^(.9){}="f" \ar@{-} "e";"f" & \downarrow & 0 ]\\
\bar{\nu}_{\tau} & [ \downarrow & {*}, & \uparrow & \downarrow ] \\
\ar@{}[rrrr]^(-.2){}="a"^(1.15){}="b" \ar@{-} "a";"b" &&&&\\
 & [ \downarrow & {*}, & \downarrow & 0 ] } 
& 
\xymatrix @C=-.3pc @R-2pc{
\bar{\mu}_{\uparrow} & [ \downarrow \ar@{}[rrrd]^(.05){}="c"^(.95){}="d" \ar@{-} "c";"d" & \uparrow, \ar@{}[rd]^(.05){}="e"^(.9){}="f" \ar@{-} "e";"f" & 0 & \downarrow ]\\
\nu_{\mu} & [ {*} & \downarrow, & \downarrow & \uparrow ] \\
\ar@{}[rrrr]^(-.2){}="a"^(1.15){}="b" \ar@{-} "a";"b" &&&&\\
 & [ {*} & \downarrow, & 0 & \downarrow ] } 
&
\xymatrix @C=-.3pc @R-2pc{
\tau_{\downarrow} & [ \uparrow \ar@{}[rrrd]^(.05){}="c"^(.95){}="d" \ar@{-} "c";"d" & \uparrow, \ar@{}[rd]^(.05){}="e"^(.9){}="f" \ar@{-} "e";"f" & \downarrow & 0 ]\\
\bar{\nu}_{\tau} & [ \downarrow & {*}, & \uparrow & \uparrow ] \\
\ar@{}[rrrr]^(-.2){}="a"^(1.15){}="b" \ar@{-} "a";"b" &&&&\\
 & [ \downarrow & {*}, & \downarrow & 0 ] } 
& 
\xymatrix @C=-.3pc @R-2pc{
\bar{\tau}_{\uparrow} & [ \uparrow \ar@{}[rrrd]^(.05){}="c"^(.95){}="d" \ar@{-} "c";"d" & \uparrow, \ar@{}[rd]^(.05){}="e"^(.9){}="f" \ar@{-} "e";"f" & 0 & \downarrow ]\\
\nu_{\tau} & [ {*} & \downarrow, & \uparrow & \uparrow ] \\
\ar@{}[rrrr]^(-.2){}="a"^(1.15){}="b" \ar@{-} "a";"b" &&&&\\
 & [ {*} & \downarrow, & 0 & \downarrow ] } 
\\
\addlinespace
\xymatrix @C=-.3pc @R-2pc{
d_{b} & \textcolor{red}{\pmb{(}} 0 & 0, & b & 0 ]\\
\bar{u}_{a} & [ b & {*}, & 0 & 0 \textcolor{red}{\pmb{)}} \\
\ar@{}[rrrr]^(-.2){}="a"^(1.15){}="b" \ar@{-} "a";"b" &&&&\\
 & [ b & {*}, & b & 0 ] } 
& 
\xymatrix @C=-.3pc @R-2pc{
\bar{d}_{a} & [ 0 & 0, & 0 & b \textcolor{red}{\pmb{)}}\\
u_{b} & \textcolor{red}{\pmb{(}} {*} & b, & 0 & 0 ] \\
\ar@{}[rrrr]^(-.2){}="a"^(1.15){}="b" \ar@{-} "a";"b" &&&&\\
 & [ {*} & b, & 0 & b ] } 
&
\xymatrix @C=-.3pc @R-2pc{
s_{b} & \textcolor{red}{\pmb{(}} a \ar@{}[rrrd]^(.05){}="c"^(.95){}="d" \ar@{-} "c";"d" & b, \ar@{}[rd]^(.05){}="e"^(.9){}="f" \ar@{-} "e";"f" & b & 0 ]\\
\bar{c}_{a} & [ b & {*}, & a & b \textcolor{red}{\pmb{)}} \\
\ar@{}[rrrr]^(-.2){}="a"^(1.15){}="b" \ar@{-} "a";"b" &&&&\\
 & [ b & {*}, & b & 0 ] } 
& 
\xymatrix @C=-.3pc @R-2pc{
\bar{s}_{a} & [ b \ar@{}[rrrd]^(.05){}="c"^(.95){}="d" \ar@{-} "c";"d" & a, \ar@{}[rd]^(.05){}="e"^(.9){}="f" \ar@{-} "e";"f" & 0 & b \textcolor{red}{\pmb{)}}\\
c_{b} & \textcolor{red}{\pmb{(}} {*} & b, & b & a ] \\
\ar@{}[rrrr]^(-.2){}="a"^(1.15){}="b" \ar@{-} "a";"b" &&&&\\
 & [ {*} & b, & 0 & b ] } 
&
\xymatrix @C=-.3pc @R-2pc{
b_{b} & \textcolor{red}{\pmb{(}} a \ar@{}[rrrd]^(.05){}="c"^(.95){}="d" \ar@{-} "c";"d" & a, \ar@{}[rd]^(.05){}="e"^(.9){}="f" \ar@{-} "e";"f" & b & 0 ]\\
\bar{t}_{a} & [ b & {*}, & a & a \textcolor{red}{\pmb{)}} \\
\ar@{}[rrrr]^(-.2){}="a"^(1.15){}="b" \ar@{-} "a";"b" &&&&\\
 & [ b & {*}, & b & 0 ] } 
& 
\xymatrix @C=-.3pc @R-2pc{
\bar{b}_{a} & [ a \ar@{}[rrrd]^(.05){}="c"^(.95){}="d" \ar@{-} "c";"d" & a, \ar@{}[rd]^(.05){}="e"^(.9){}="f" \ar@{-} "e";"f" & 0 & b \textcolor{red}{\pmb{)}} \\
t_{b} & \textcolor{red}{\pmb{(}} {*} & b, & a & a ] \\
\ar@{}[rrrr]^(-.2){}="a"^(1.15){}="b" \ar@{-} "a";"b" &&&&\\
 & [ {*} & b, & 0 & b ] } 
\end{array}
$}}
}\end{equation*}
with $a, b \in \{ \uparrow, \downarrow \}$, $a \not = b$, produce configurations that violate the Pauli exclusion principle, and thus are not geoms.
Similarly, the concatenations
\begin{equation*}
{\arraycolsep=2.2pt
\text{\footnotesize{$
\begin{array}{cccccc} 
\xymatrix @C=-.3pc @R-2pc{
d_{\downarrow} & \textcolor{red}{\pmb{(}} 0 & 0, & \downarrow & 0 ]\\
\bar{u}_{\downarrow} & [ \uparrow & {*}, & 0 & 0 \textcolor{red}{\pmb{)}} \\
\ar@{}[rrrr]^(-.2){}="a"^(1.15){}="b" \ar@{-} "a";"b" &&&&\\
 & [ \uparrow & {*}, & \downarrow & 0 ] } 
& 
\xymatrix @C=-.3pc @R-2pc{
\bar{d}_{\uparrow} & [ 0 & 0, & 0 & \downarrow \! \textcolor{red}{\pmb{)}} \\
u_{\uparrow} & \textcolor{red}{\pmb{(}} {*} & \uparrow, & 0 & 0 ] \\
\ar@{}[rrrr]^(-.2){}="a"^(1.15){}="b" \ar@{-} "a";"b" &&&&\\
 & [ {*} & \uparrow, & 0 & \downarrow ] } 
&
\xymatrix @C=-.3pc @R-2pc{
s_{\downarrow} & \textcolor{red}{\pmb{(}} \! \uparrow \ar@{}[rrrd]^(.05){}="c"^(.95){}="d" \ar@{-} "c";"d" & \downarrow \ar@{}[rd]^(.05){}="e"^(.9){}="f" \ar@{-} "e";"f" & \downarrow & 0 ]\\
\bar{c}_{\downarrow} & [ \uparrow & {*}, & \downarrow & \uparrow \! \textcolor{red}{\pmb{)}}\\
\ar@{}[rrrr]^(-.2){}="a"^(1.15){}="b" \ar@{-} "a";"b" &&&&\\
 & [ \uparrow & {*}, & \downarrow & 0 ] } 
& 
\xymatrix @C=-.3pc @R-2pc{
\bar{s}_{\uparrow} & [ \downarrow \ar@{}[rrrd]^(.05){}="c"^(.95){}="d" \ar@{-} "c";"d" & \uparrow \ar@{}[rd]^(.05){}="e"^(.9){}="f" \ar@{-} "e";"f" & 0 & \downarrow \! \textcolor{red}{\pmb{)}}\\
c_{\uparrow} & \textcolor{red}{\pmb{(}} {*} & \uparrow, & \uparrow & \downarrow ] \\
\ar@{}[rrrr]^(-.2){}="a"^(1.15){}="b" \ar@{-} "a";"b" &&&&\\
 & [ {*} & \uparrow, & 0 & \uparrow ] } 
&
\xymatrix @C=-.3pc @R-2pc{
b_{\downarrow} & \textcolor{red}{\pmb{(}} \! \uparrow \ar@{}[rrrd]^(.05){}="c"^(.95){}="d" \ar@{-} "c";"d" & \uparrow, \ar@{}[rd]^(.05){}="e"^(.9){}="f" \ar@{-} "e";"f" & \downarrow & 0 ]\\
\bar{t}_{\downarrow} & [ \uparrow & {*}, & \downarrow & \downarrow \! \textcolor{red}{\pmb{)}}\\
\ar@{}[rrrr]^(-.2){}="a"^(1.15){}="b" \ar@{-} "a";"b" &&&&\\
 & [ \uparrow & {*}, & \downarrow & 0 ] } 
& 
\xymatrix @C=-.3pc @R-2pc{
\bar{b}_{\uparrow} & [ \uparrow \ar@{}[rrrd]^(.05){}="c"^(.95){}="d" \ar@{-} "c";"d" & \uparrow, \ar@{}[rd]^(.05){}="e"^(.9){}="f" \ar@{-} "e";"f" & 0 & \downarrow \! \textcolor{red}{\pmb{)}}\\
t_{\uparrow} & \textcolor{red}{\pmb{(}} {*} & \uparrow, & \downarrow & \downarrow ] \\
\ar@{}[rrrr]^(-.2){}="a"^(1.15){}="b" \ar@{-} "a";"b" &&&&\\
 & [ {*} & \uparrow, & 0 & \downarrow ] } 
\end{array}
$}}
}\end{equation*}
also produce configurations that are not geoms, since ${[\uparrow \! {*}]}$ and ${[{*} \! \uparrow]}$ are not orbitals by Theorem \ref{allorbitals}.
Consequently, these twenty-one concatenations are not fusions, and so are excluded.\footnote{In particular, our model does not admit sterile neutrinos, that is, right-handed neutrinos and left-handed anti-neutrinos that do not interact with the $W^{\pm}$ geoms.}
\end{proof}

\section{Differences with the standard model interactions} \label{differences}

Our model admits two new interaction vertices, $\gamma_{\uparrow}\gamma_{\downarrow}Z_0$ and $Z_0Z_0Z_0$ given in Table \ref{WWgamma}, that do not occur in the standard model.
These are the only two fusions that do not correspond to standard model vertices.
It is possible that they may be replacements for the four-boson vertices of the standard model, 
\begin{center}
\begin{tikzpicture}
  \begin{feynman}
    \vertex (1);
    \vertex [above left=of 1] (2) {\( W^+ \)};
    \vertex [below left=of 1] (3)  {\( W^- \)};
    \vertex [above right=of 1] (4) {\( \gamma/W^+/Z\)};
    \vertex [below right=of 1] (5) {\( \gamma/W^-/Z\)};
    \diagram* {
      (1) -- [charged boson] (2),
      (3) -- [charged boson](1),
      (1) -- [boson] (4),
      (1) -- [boson] (5),
    };
  \end{feynman}
\end{tikzpicture}
\quad \quad\raisebox{8ex}{$\mapsto$}\quad \quad
\begin{tikzpicture}
  \begin{feynman}
  \vertex (1);
    \vertex [above left=of 1] (2) {\( W^+_0 \)};
    \vertex [below left=of 1] (3)  {\( W^-_0 \)};
    \vertex [right=of 1] (6);
    \vertex [above right=of 6] (4) {\( \gamma_{\uparrow}/W^+_0/Z_0\)};
    \vertex [below right=of 6] (5) {\( \gamma_{\downarrow}/W^-_0/Z_0\)};
    \diagram* {
      (1) -- [charged boson] (2), (1) -- [boson, edge label=\( Z_0 \)] (6),
      (3) -- [charged boson](1),
      (6) -- [boson] (4),
      (6) -- [boson] (5),
    };
  \end{feynman}
\end{tikzpicture}
\end{center}
and $Z_0Z_0Z_0$ may be a replacement for the three-gluon vertex $ggg$,
\begin{center}
\begin{tikzpicture}
  \begin{feynman}
    \vertex (2);
    \vertex [above left=of 2] (1) {\(g \)};
    \vertex [right=of 2] (3) {\(g \)};
    \vertex [below left=of 2] (4) {\(g \)};
    \diagram* {
      (1) -- [gluon] (2) -- [gluon] (3),
      (4) -- [gluon] (2),
    };
  \end{feynman}
\end{tikzpicture}
\quad \quad\raisebox{8ex}{$\mapsto$}\quad \quad
\begin{tikzpicture}
  \begin{feynman}
    \vertex (2);
    \vertex [above left=of 2] (1) {\(Z_0 \)};
    \vertex [right=of 2] (3) {\(Z_0 \)};
    \vertex [below left=of 2] (4) {\(Z_0 \)};
    \diagram* {
      (1) -- [boson] (2) -- [boson] (3),
      (4) -- [boson] (2),
    };
  \end{feynman}
\end{tikzpicture}
\end{center}
Note that the trivalent $Z$-boson vertex only occurs with the longitudinal polarization $Z_0 = {[\uparrow \downarrow, \downarrow \uparrow]}$ of the $Z$ boson.
Moreover, in QCD there is an infrared suppression of the three-gluon vertex, and at energies where the three-gluon vertex occurs gluons acquire a dynamical mass scale (see \cite{PAF} and references therein). 
This is qualitatively consistent with our replacement $ggg \mapsto Z_0Z_0Z_0$, since the $Z_0$ boson is massive. 

In addition, due to restrictions on the allowable spin and polarization states from fusions, our model does not admit all Feynman diagrams that arise in the standard model. 
For example, the diagrams
\begin{center}
\begin{tabular}{ccc}
\raisebox{7.5ex}{(\textsc{a})} \ \ \ \
\begin{tikzpicture}
  \begin{feynman}
    \vertex (1) {\(Z_{\uparrow} \)};
    \vertex [right=of 1] (2);
    \vertex [right=of 2] (3);
    \vertex [right=of 3] (4) {\( \gamma \)};
    \diagram* {
      (1) -- [boson] (2) -- [fermion, half left, edge label=\(e_{\uparrow}\)] (3) -- [boson] (4),
      (3) -- [fermion, half left, edge label=\( \bar{e}_{\downarrow} \)] (2),
    };
  \end{feynman}
\end{tikzpicture}
&
\ \ \ \ \ \ 
&
\raisebox{7.5ex}{(\textsc{b})} \ \ \ \
\begin{tikzpicture}
  \begin{feynman}
    \vertex (1) {\( H \)};
    \vertex [right=of 1] (2);
    \vertex [above right=of 2] (3);
    \vertex [below right=of 2] (4);
    \vertex [right=of 3] (5) {\( \gamma \)};
    \vertex [right=of 4] (6) {\( \gamma \)};
    \diagram* {
      (1) -- [scalar] (2) -- [charged boson, edge label=\( W^- \)] (3) -- [charged boson, edge label=\( W^- \)] (4) -- [charged boson, edge label=\( W^+ \)] (2),
      (3) -- [boson] (5), (4) -- [boson] (6),
    };
  \end{feynman}
\end{tikzpicture}
\\
\multicolumn{3}{c}{
\raisebox{7.5ex}{(\textsc{c})} \ \ \ \ 
\begin{tikzpicture}
  \begin{feynman}
    \vertex (0) {\( H \)};
    \vertex [right=of 0] (1);
    \vertex [right=of 1] (2);
    \vertex [right=of 2] (3);
    \vertex [right=of 3] (4) {\(e_{\uparrow} \)};
    \vertex [below right=of 3] (5) {\( \bar{\nu}_e \)};
    \vertex [below right=of 2] (6) {\(W^+_{\downarrow} \)};
    \vertex [below right=of 1] (7) {\( Z_{\downarrow} \)};
    \diagram* {
      (0) -- [scalar] (1) -- [boson, edge label=\( Z_{\uparrow} \)] (2) -- [charged boson, edge label=\( W^-_{\uparrow} \)] (3) -- [fermion] (4),
      (3) -- [anti fermion] (5), (6) -- [charged boson] (2), (1) -- [boson] (7),
    };
  \end{feynman}
\end{tikzpicture}}
\end{tabular}
\end{center}
do not occur in our model.
However, although our model does not admit (\textsc{b}) (see e.g.\ \cite[Figure 1]{CMS}), it does admit the Higgs decay $H \to \gamma Z$ through the $W$-boson, as well as $H \to \gamma \gamma$ through any charged lepton $\ell$ or down-type quark $q$:
\begin{center}
\begin{tikzpicture}
  \begin{feynman}
    \vertex (1) {\( H \)};
    \vertex [right=of 1] (2);
    \vertex [above right=of 2] (3);
    \vertex [below right=of 2] (4);
    \vertex [right=of 3] (5) {\( \gamma_a \)};
    \vertex [right=of 4] (6) {\( Z_a \)};
    \diagram* {
      (1) -- [scalar] (2) -- [charged boson, edge label=\( W^-_a \)] (3) -- [charged boson, edge label=\( W^-_a \)] (4) -- [charged boson, edge label=\( W^+_b \)] (2),
      (3) -- [boson] (5), (4) -- [boson] (6),
    };
  \end{feynman}
\end{tikzpicture}
\ \ \ \ \ \ 
\begin{tikzpicture}
  \begin{feynman}
    \vertex (1) {\( H \)};
    \vertex [right=of 1] (2);
    \vertex [above right=of 2] (3);
    \vertex [below right=of 2] (4);
    \vertex [right=of 3] (5) {\( \gamma_a \)};
    \vertex [right=of 4] (6) {\( \gamma_a \)};
    \diagram* {
      (1) -- [scalar] (2) -- [fermion, edge label=\( \ell_a/q_a \)] (3) -- [fermion, edge label=\( \ell_a/q_a \)] (4) -- [fermion, edge label=\( \bar{\ell}_a/\bar{q}_a \)] (2),
      (3) -- [boson] (5), (4) -- [boson] (6),
    };
  \end{feynman}
\end{tikzpicture}
\end{center}
with $a,b \in \{ \uparrow, \downarrow \}$, $a \not = b$.
Our model also admits the diagram
\begin{center}
\begin{tikzpicture}
  \begin{feynman}
    \vertex (0) {\( H \)};
    \vertex [right=of 0] (1);
    \vertex [above right=of 1] (2);
    \vertex [right=of 2] (3);
    \vertex [below right=of 1] (4);
    \vertex[right=of 4] (6);
    \vertex [right=of 3] (5) {\( \gamma_a \)};
    \vertex [right=of 6] (7) {\( \gamma_a \)};
    \diagram* {
      (0) -- [scalar] (1) -- [charged boson, edge label=\( W^-_0 \)] (2) -- [boson, edge label=\( Z_0 \)] (3) -- [boson] (5),
      (2) -- [charged boson, edge label=\( W^-_0 \)] (4) -- [charged boson, edge label=\( W_0^+ \)] (1), 
      (4) -- [boson, edge label=\( Z_0 \), swap] (6) -- [boson] (7),
      (3) -- [boson, edge label=\( \gamma_b \)] (6),
    };
  \end{feynman}
\end{tikzpicture}
\end{center}
using the new $\gamma_{\uparrow}\gamma_{\downarrow}Z_0$ vertex.
Furthermore, although our model does not admit (\textsc{c}), it does admit the similar diagrams
\begin{center}
\begin{tabular}{ccc}
\begin{tikzpicture}
  \begin{feynman}
    \vertex (0) {\( H \)};
    \vertex [right=of 0] (1);
    \vertex [right=of 1] (2);
    \vertex [right=of 2] (3) {\(e_{\uparrow} \)};
    \vertex [below right=of 2] (4) {\( \bar{\nu}_e \)};
    \vertex [below right=of 1] (5) {\(W^+_0 \)};
    \diagram* {
      (0) -- [scalar] (1) -- [charged boson, edge label=\( W^-_0 \)] (2) -- [fermion] (3),
      (2) -- [anti fermion] (4), (5) -- [charged boson] (1),
    };
  \end{feynman}
\end{tikzpicture}
& \ \ \ &
\begin{tikzpicture}
  \begin{feynman}
    \vertex (0) {\( H \)};
    \vertex [right=of 0] (1);
    \vertex [right=of 1] (2);
    \vertex [right=of 2] (3) {\(e_{\uparrow} \)};
    \vertex [below right=of 2] (4) {\( \bar{e} \)};
    \vertex [below right=of 1] (5) {\(Z_{\downarrow} \)};
    \diagram* {
      (0) -- [scalar] (1) -- [charged boson, edge label=\( Z_{\uparrow} \)] (2) -- [fermion] (3),
      (2) -- [anti fermion] (4), (5) -- [charged boson] (1),
    };
  \end{feynman}
\end{tikzpicture}
\\
\multicolumn{3}{c}{
\begin{tikzpicture}
  \begin{feynman}
    \vertex (0) {\( Z_{0} \)};
    \vertex [right=of 0] (1);
    \vertex [right=of 1] (2);
    \vertex [right=of 2] (3) {\(e_{\uparrow} \)};
    \vertex [below right=of 2] (4) {\( \bar{\nu}_e \)};
    \vertex [below right=of 1] (5) {\(W^+_{0} \)};
    \diagram* {
      (0) -- [boson] (1) -- [charged boson, edge label=\( W^-_{0} \)] (2) -- [fermion] (3),
      (2) -- [anti fermion] (4), (5) -- [charged boson] (1),
    };
  \end{feynman}
\end{tikzpicture}}
\end{tabular}
\end{center}

We hope that further investigation of these and similar Feynman diagram discrepancies will yield testable predictions of our model that disagree with standard model predictions. 

\section{Interaction vertices of the predicted new boson}

Our composite model predicts the existence of precisely five new bosonic polarization states, denoted $x_0$ and $x_{ab}$, $a,b \in \{ \uparrow, \downarrow \}$, given in Table \ref{particles}.
If these states are massive, then they form a polarization basis for a spin-$2$ boson $x$.
The predicted interactions of this boson, determined by Definition \ref{fusion def}, are given in Table \ref{newinteractions}.

We have only imposed two conditions in constructing our composite model, namely, a Dirac equation constraint (\ref{coupling constraints}) and the Pauli exclusion principle; see \cite[Theorem 7.1]{B1}. 
However, by imposing further conditions we may exclude $x_0$ and $x_{ab}$ from our model without effecting the standard model geoms. 
Indeed, if we also assume that a geom may have at most one $*$-orbital, then $x_0$ could not form.
If we assume that an orbital needs to be `full', in the sense that both of its components are nonzero, before a second orbital can be nonempty, then the states $x_{ab}$ could not form.
This would be similar to the Aufbau principle, which asserts that the electrons in an atom fill lower energy orbitals before filling higher energy ones.

\setlength{\tabcolsep}{2.2pt}
\begin{longtable}[h]{cccccc}
\caption{Interaction vertices of the predicted spin-$2$ $x$ boson.
Here, $a,b,c,d \in \{ \uparrow, \downarrow \}$ with $a \not = b$, $c \not = d$.}
\endfirsthead  
\multicolumn{6}{@{}l}{(\emph{table continued from previous page})}
\endhead
\multicolumn{6}{@{}l}{(\emph{table continued on next page})}
\endfoot
\endlastfoot  
{\footnotesize $\xymatrix @C=-.3pc @R-2pc{
e_{a} & [ 0 & 0, & a & 0 ]\\
\bar{e}_{c} & [ 0 & d, & 0 & 0 ] \\
\ar@{}[rrrr]^(-.2){}="a"^(1.15){}="b" \ar@{-} "a";"b" &&&&\\
x_{ad} & [ 0 & d, & a & 0 ] } $}
& 
{\footnotesize $\xymatrix @C=-.3pc @R-2pc{
\bar{\nu}_e & [ 0 & 0, & \downarrow & {*} \ar@{}[llld]^(.05){}="e"^(.95){}="f" \ar@{-} "e";"f" ]\\
\nu_{e} & [ {*} & \downarrow, & 0 & 0 ] \\
\ar@{}[rrrr]^(-.2){}="a"^(1.15){}="b" \ar@{-} "a";"b" &&&&\\
x_{ad} & [ 0 & d, & a & 0 ] } $}
&
{\footnotesize $\xymatrix @C=-.3pc @R-2pc{
\mu_a & [ b \ar@{}[rrrd]^(.05){}="c"^(.95){}="d" \ar@{-} "c";"d" & a, \ar@{}[rd]^(.05){}="e"^(.9){}="f" \ar@{-} "e";"f" & a & 0 ]\\
\bar{\mu}_c & [ 0 & d, & d & c ] \\
\ar@{}[rrrr]^(-.2){}="a"^(1.15){}="b" \ar@{-} "a";"b" &&&&\\
x_{ad} & [ 0 & d, & a & 0 ] } $}
& 
&
{\footnotesize $\xymatrix @C=-.3pc @R-2pc{
\tau_{a} & [ b \ar@{}[rrrd]^(.05){}="c"^(.95){}="d" \ar@{-} "c";"d" & b, \ar@{}[rd]^(.05){}="e"^(.9){}="f" \ar@{-} "e";"f" & a & 0 ]\\
\bar{\tau}_c & [ 0 & d, & c & c ] \\
\ar@{}[rrrr]^(-.2){}="a"^(1.15){}="b" \ar@{-} "a";"b" &&&&\\
x_{ad} & [ 0 & d, & a & 0 ] } $}
& 
\\
\addlinespace
& 
{\footnotesize $\xymatrix @C=-.3pc @R-2pc{
\bar{\nu}_e & [ 0 & 0, & \downarrow & {*} ]\\
\nu_{e} & [ {*} & \downarrow, & 0 & 0 ] \\
\ar@{}[rrrr]^(-.2){}="a"^(1.15){}="b" \ar@{-} "a";"b" &&&&\\
x_0 & [ {*} & \downarrow, & \downarrow & {*} ] } $}
&
& 
{\footnotesize $\xymatrix @C=-.3pc @R-2pc{
\bar{\nu}_{\mu} & [ \uparrow \ar@{}[rrrd]^(.05){}="c"^(.95){}="d" \ar@{-} "c";"d" & \downarrow, \ar@{}[rd]^(.05){}="e"^(.9){}="f" \ar@{-} "e";"f" & \downarrow & {*} ]\\
\nu_{\mu} & [ {*} & \downarrow, & \downarrow & \uparrow ] \\
\ar@{}[rrrr]^(-.2){}="a"^(1.15){}="b" \ar@{-} "a";"b" &&&&\\
x_0 & [ {*} & \downarrow, & \downarrow & {*} ] } $}
&
& 
{\footnotesize $\xymatrix @C=-.3pc @R-2pc{
\bar{\nu}_{\tau} & [ \uparrow \ar@{}[rrrd]^(.05){}="c"^(.95){}="d" \ar@{-} "c";"d" & \uparrow, \ar@{}[rd]^(.05){}="e"^(.9){}="f" \ar@{-} "e";"f" & \downarrow & {*} ]\\
\nu_{\tau} & [ {*} & \downarrow, & \uparrow & \uparrow ] \\
\ar@{}[rrrr]^(-.2){}="a"^(1.15){}="b" \ar@{-} "a";"b" &&&&\\
x_0 & [ {*} & \downarrow, & \downarrow & {*} ] } $}
\\
\addlinespace
{\footnotesize $\xymatrix @C=-.3pc @R-2pc{
d_{a} & \textcolor{red}{\pmb{(}} 0 & 0, & a & 0 ]\\
\bar{d}_{c} & [ 0 & d, & 0 & 0 \textcolor{red}{\pmb{)}} \\
\ar@{}[rrrr]^(-.2){}="a"^(1.15){}="b" \ar@{-} "a";"b" &&&&\\
x_{ad} & [ 0 & d, & a & 0 ] } $}
& 
{\footnotesize $\xymatrix @C=-.3pc @R-2pc{
\bar{u}_{a} & [ 0 & 0, & a & {*} \ar@{}[llld]^(.05){}="e"^(.95){}="f" \ar@{-} "e";"f" \textcolor{red}{\pmb{)}}\\
u_{c} & \textcolor{red}{\pmb{(}} {*} & d, & 0 & 0 ] \\
\ar@{}[rrrr]^(-.2){}="a"^(1.15){}="b" \ar@{-} "a";"b" &&&&\\
x_{ad} & [ 0 & a, & d & 0 ] } $}
&
{\footnotesize $\xymatrix @C=-.3pc @R-2pc{
s_a & \textcolor{red}{\pmb{(}} b \ar@{}[rrrd]^(.05){}="c"^(.95){}="d" \ar@{-} "c";"d" & a, \ar@{}[rd]^(.05){}="e"^(.9){}="f" \ar@{-} "e";"f" & a & 0 ]\\
\bar{s}_c & [ 0 & d, & d & c \textcolor{red}{\pmb{)}} \\
\ar@{}[rrrr]^(-.2){}="a"^(1.15){}="b" \ar@{-} "a";"b" &&&&\\
x_{ad} & [ 0 & d, & a & 0 ] } $}
& 
&
{\footnotesize $\xymatrix @C=-.3pc @R-2pc{
b_{a} & \textcolor{red}{\pmb{(}} b \ar@{}[rrrd]^(.05){}="c"^(.95){}="d" \ar@{-} "c";"d" & b, \ar@{}[rd]^(.05){}="e"^(.9){}="f" \ar@{-} "e";"f" & a & 0 ]\\
\bar{b}_c & [ 0 & d, & c & c \textcolor{red}{\pmb{)}} \\
\ar@{}[rrrr]^(-.2){}="a"^(1.15){}="b" \ar@{-} "a";"b" &&&&\\
x_{ad} & [ 0 & d, & a & 0 ] } $}
& 
\\
\addlinespace
& 
{\footnotesize $\xymatrix @C=-.3pc @R-2pc{
\bar{u}_{\downarrow} & [ 0 & 0, & \downarrow & {*} \textcolor{red}{\pmb{)}}\\
u_{\uparrow} & \textcolor{red}{\pmb{(}} {*} & \downarrow, & 0 & 0 ] \\
\ar@{}[rrrr]^(-.2){}="a"^(1.15){}="b" \ar@{-} "a";"b" &&&&\\
x_0 & [ {*} & \downarrow, & \downarrow & {*} ] } $}
&
& 
{\footnotesize $\xymatrix @C=-.3pc @R-2pc{
\bar{c}_{\downarrow} & [ \uparrow \ar@{}[rrrd]^(.05){}="c"^(.95){}="d" \ar@{-} "c";"d" & \downarrow, \ar@{}[rd]^(.05){}="e"^(.9){}="f" \ar@{-} "e";"f" & \downarrow & {*} \textcolor{red}{\pmb{)}}\\
c_{\uparrow} & \textcolor{red}{\pmb{(}} {*} & \downarrow, & \downarrow & \uparrow ] \\
\ar@{}[rrrr]^(-.2){}="a"^(1.15){}="b" \ar@{-} "a";"b" &&&&\\
x_0 & [ {*} & \downarrow, & \downarrow & {*} ] } $}
&
& 
{\footnotesize $\xymatrix @C=-.3pc @R-2pc{
\bar{t}_{\downarrow} & [ \uparrow \ar@{}[rrrd]^(.05){}="c"^(.95){}="d" \ar@{-} "c";"d" & \uparrow, \ar@{}[rd]^(.05){}="e"^(.9){}="f" \ar@{-} "e";"f" & \downarrow & {*} \textcolor{red}{\pmb{)}}\\
t_{\uparrow} & \textcolor{red}{\pmb{(}} {*} & \downarrow, & \uparrow & \uparrow ] \\
\ar@{}[rrrr]^(-.2){}="a"^(1.15){}="b" \ar@{-} "a";"b" &&&&\\
x_0 & [ {*} & \downarrow, & \downarrow & {*} ] } $}
\\
\addlinespace
{\footnotesize $\xymatrix @C=-.3pc @R-2pc{
\gamma_{a} & [ 0 & 0, & a & b \ar@{}[llld]^(.05){}="e"^(.95){}="f" \ar@{-} "e";"f" ]\\
\gamma_{c} & [ c & d, & 0 & 0 ] \\
\ar@{}[rrrr]^(-.2){}="a"^(1.15){}="b" \ar@{-} "a";"b" &&&&\\
x_{ad} & [ 0 & d, & a & 0 ] } $}
& 
{\footnotesize $\xymatrix @C=-.3pc @R-2pc{
\gamma_{a} & [ 0 & 0, & a \ar@{}[ld]^(.11){}="e"^(.95){}="f" \ar@{-} "e";"f" & b ]\\
\gamma_{c} & [ c & d, & 0 & 0 ] \\
\ar@{}[rrrr]^(-.2){}="a"^(1.15){}="b" \ar@{-} "a";"b" &&&&\\
x_{cb} & [ c & 0, & 0 & b ] } $}
& 
{\footnotesize $\xymatrix @C=-.3pc @R-2pc{
Z_a & [ a & a, \ar@{}[rd]^(.05){}="e"^(.9){}="f" \ar@{-} "e";"f" & 0 & 0 ]\\
Z_c & [ 0 & 0, & c & c ] \\
\ar@{}[rrrr]^(-.2){}="a"^(1.15){}="b" \ar@{-} "a";"b" &&&&\\
x_{ac} & [ a & 0, & 0 & c ] } $}
&
{\footnotesize $\xymatrix @C=-.3pc @R-2pc{
W^-_0 & [ \downarrow \ar@{}[rrrd]^(.05){}="c"^(.95){}="d" \ar@{-} "c";"d" & {*}, \ar@{}[rd]^(.05){}="e"^(.9){}="f" \ar@{-} "e";"f" & \uparrow & 0 ]\\
W^+_0 & [ 0 & \uparrow, & {*} & \downarrow ] \\
\ar@{}[rrrr]^(-.2){}="a"^(1.15){}="b" \ar@{-} "a";"b" &&&&\\
x_{\uparrow \uparrow} & [ 0 & \uparrow, & \uparrow & 0 ] } $}
&
{\footnotesize $\xymatrix @C=-.3pc @R-2pc{
W^-_a & [ {*} \ar@{}[rrrd]^(.05){}="c"^(.95){}="d" \ar@{-} "c";"d" & \downarrow, \ar@{}[rd]^(.05){}="e"^(.9){}="f" \ar@{-} "e";"f" & a & 0 ]\\
W^+_c & [ 0 & d, & \downarrow & {*} ] \\
\ar@{}[rrrr]^(-.2){}="a"^(1.15){}="b" \ar@{-} "a";"b" &&&&\\
x_{ad} & [ 0 & d, & a & 0 ] } $}
&
\\
\addlinespace
&
&
&
{\footnotesize $\xymatrix @C=-.3pc @R-2pc{
W^-_0 & [ \downarrow & {*}, & \uparrow \ar@{}[ld]^(.11){}="e"^(.95){}="f" \ar@{-} "e";"f" & 0 ]\\
W^+_0 & [ 0 & \uparrow, & {*} & \downarrow ] \\
\ar@{}[rrrr]^(-.2){}="a"^(1.15){}="b" \ar@{-} "a";"b" &&&&\\
x_{0} & [ \downarrow & {*}, & {*} & \downarrow ] } $}
& 
{\footnotesize $\xymatrix @C=-.3pc @R-2pc{
W^-_a & [ {*} & \downarrow, & a \ar@{}[ld]^(.11){}="e"^(.95){}="f" \ar@{-} "e";"f" & 0 ]\\
W^+_c & [ 0 & d, & \downarrow & {*} ] \\
\ar@{}[rrrr]^(-.2){}="a"^(1.15){}="b" \ar@{-} "a";"b" &&&&\\
x_{0} & [ {*} & \downarrow, & \downarrow & {*} ] } $} 
& 
\\
\addlinespace
{\footnotesize $\xymatrix @C=-.3pc @R-2pc{
x_{aa} & [ a & 0, & 0 & a ]\\
x_{bb} & [ 0 & b, & b & 0 ] \\
\ar@{}[rrrr]^(-.2){}="a"^(1.15){}="b" \ar@{-} "a";"b" &&&&\\
Z_0 & [ a & b, & b & a ] } $}
& 
{\footnotesize $\xymatrix @C=-.3pc @R-2pc{
x_{ab} & [ a & 0, & 0 & b ]\\
x_{ba} & [ 0 & a, & b & 0 ] \\
\ar@{}[rrrr]^(-.2){}="a"^(1.15){}="b" \ar@{-} "a";"b" &&&&\\
H & [ a & a, & b & b ] } $}
&
{\footnotesize $\xymatrix @C=-.3pc @R-2pc{
x_{ab} & [ a & 0, & 0 & b \ar@{}[llld]^(.05){}="e"^(.95){}="f" \ar@{-} "e";"f" ]\\
x_{cd} & [ c & 0, & 0 & d ] \\
\ar@{}[rrrr]^(-.2){}="a"^(1.15){}="b" \ar@{-} "a";"b" &&&&\\
x_{ad} & [ a & 0, & 0 & d ] } $}
&
{\footnotesize $\xymatrix @C=-.3pc @R-2pc{
x_0 & [ {*} & \downarrow, & \downarrow \ar@{}[ld]^(.11){}="e"^(.95){}="f" \ar@{-} "e";"f" & {*} \ar@{}[llld]^(.05){}="e"^(.95){}="f" \ar@{-} "e";"f" ]\\
x_0 & [ {*} & \downarrow, & \downarrow & {*} ]\\
\ar@{}[rrrr]^(-.2){}="a"^(1.15){}="b" \ar@{-} "a";"b" &&&&\\
x_0 & [ {*} & \downarrow, & \downarrow & {*} ] } $}
&
&
\label{newinteractions}
\end{longtable} 

\ \\
\textbf{Acknowledgments.}
The author was supported by the Austrian Science Fund (FWF) grant P 34854.

\bibliographystyle{hep}
\def\cprime{$'$} \def\cprime{$'$}

\end{document}